\documentclass[prd,aps,amsfonts,amsmath, nofootinbib]{revtex4}

\input{epsf}
\usepackage{graphicx}
\usepackage{subeqnarray}

\def\rmd{{\rm d}}
\def\lsim{\mathrel{\rlap{\lower4pt\hbox{\hskip1pt$\sim$}}\raise1pt\hbox{$<$}}}
\def\gsim{\mathrel{\rlap{\lower4pt\hbox{\hskip1pt$\sim$}}\raise1pt\hbox{$>$}}}
\def\be{\begin{equation}}
\def\en{\end{equation}}
\def\bea{\begin{eqnarray}}
\def\ena{\end{eqnarray}}

\def\bx{{\bf x}}
\def\bxd{{\bf x'}}
\newcommand{\erf}[1]{(\ref{#1})}

\begin{document}


\title{``Kludge'' gravitational waveforms for a test-body orbiting a
Kerr black hole}

\author{Stanislav Babak}
\email[]{stba@aei.mpg.de}
\affiliation{Max-Planck-Institut fuer Gravitationsphysik, Albert-Einstein-Institut, Am Muchlenberg 1, D-14476 Golm bei
Potsdam, Germany}
\author{Hua Fang}
\email[]{hua@tapir.caltech.edu}
\affiliation{Theoretical Astrophysics, California Institute of Technology, Pasadena, CA 91125, USA}
\author{Jonathan R. Gair}
\email[]{jgair@ast.cam.ac.uk}
\affiliation{Institute of Astronomy, Madingley Road, Cambridge, CB3 0HA, UK}
\author{Kostas Glampedakis}
\email[]{kg1@maths.soton.ac.uk}
\affiliation{School of Mathematics, University of Southampton, Southampton SO17 1BJ, UK}
\author{Scott A. Hughes}
\email[]{sahughes@mit.edu}
\affiliation{Department of Physics and MIT Kavli Institute,
Massachusetts Institute of Technology, 77 Massachusetts Avenue,
Cambridge, MA 02139, USA}
\begin{abstract}
One of the most exciting potential sources of gravitational waves for
low-frequency, space-based gravitational wave (GW) detectors such as
the proposed Laser Interferometer Space Antenna (LISA) is the inspiral 
of compact objects into
massive black holes in the centers of galaxies.  The detection of
waves from such ``extreme mass ratio inspiral'' systems (EMRIs) and
extraction of information from those waves require template waveforms.
The systems' extreme mass ratio means that their waveforms can be
determined accurately using black hole perturbation theory.  Such
calculations are computationally very expensive. There is a pressing
need for families of approximate waveforms that may be generated
cheaply and quickly but which still capture the main features of true
waveforms.  In this paper, we introduce a family of such ``kludge''
waveforms and describe ways to generate them.  Different kinds of
``kludges'' have already been used to scope out data analysis issues
for LISA.  The models we study here are based on computing a particle's
inspiral trajectory in Boyer-Lindquist coordinates, and subsequent
identification of these coordinates with flat space spherical polar
coordinates. A gravitational waveform may then be computed from the
multipole moments of the trajectory in these coordinates, using well
known solutions of the linearised gravitational perturbation equations
in flat spacetime.  We compute waveforms using a standard slow-motion
quadrupole formula, a quadrupole/octupole formula, and a fast-motion,
weak-field formula originally developed by Press.  We assess these
approximations by comparing to accurate waveforms obtained by solving
the Teukolsky equation in the adiabatic limit (neglecting GW
backreaction).  We find that the kludge waveforms do extremely well at
approximating the true gravitational waveform, having overlaps with
the Teukolsky waveforms of $95\% $ or higher over most of the
parameter space for which comparisons can currently be made.  Indeed,
we find these kludges to be of such high quality (despite their ease
of calculation) that it is possible they may play some role in the
final search of LISA data for EMRIs.
\end{abstract}

\maketitle


\section{Introduction-Motivation}

The proposed Laser Interferometer Space Antenna (LISA)~\cite{LISA}, is
expected to provide a variety of high-precision gravitational wave
(GW) measurements.  One of the most interesting targets for this
space-based detector are the GWs generated by stellar-mass compact
objects inspiralling into (super)massive black holes
[(S)MBHs\footnote{LISA's sensitivity is primarily to events involving
black holes on the low end of the mass spectrum seen in many galaxies
--- around $10^5$ to (a few)$\times 10^7\,M_\odot$. We emphasize this
point because the name ``supermassive black hole'' is often taken to
refer to a hole with mass in the range $10^7 - 10^9\,M_\odot$.  GW
events from such black holes will typically be at frequencies too low
for LISA's sensitivity.}]. Accumulated astrometric observations
provide strong support in favor of the existence of a ``dark'' compact
object in the core of every galaxy (for which the central parsec
region can be resolved)~\cite{obs1}. With masses ranging between
$10^6-10^9 M_{\odot}$ these objects are believed to be massive Kerr
black holes~\cite{MBH1}. It is also believed that these holes are the
``quiet'' remnants of an older quasar population~\cite{MBH_review}.

Multi-body interactions in the ``cusp'' stellar population surrounding
these SMBHs can put stellar mass compact objects onto orbits that come
close to the central black hole. If the object passes sufficiently
close to the (S)MBH, it may be captured and subsequently spiral in by
the emission of GWs \cite{sigurdsson,rees1,rees2}. Initially, the
captured bodies are expected to be on ``generic'' orbits, i.e.,
eccentric (with eccentricity $e \approx 1$) and inclined with respect
to the central black hole's equatorial plane
\cite{sigurdsson,freitag}. These orbits evolve adiabatically due to GW
emission, decreasing in eccentricity and periastron while the
inclination of the orbit remains approximately constant, but increases
slightly (see~\cite{GHK,GHK2,kg_review} for an approximate description
of the full inspiral).

For central black hole masses in the range $10^5 M_{\odot} \lsim M
\lsim 10^7 M_{\odot}$, the GWs emitted during the inspiral will be at
frequencies close to the floor of the LISA noise curve ($\sim 3$
mHz). LISA will detect the bursts of radiation emitted near periapse
throughout the inspiral, but these will not be individually
resolvable~\cite{leor04} unless they occur in our
Galaxy~\cite{Rubbo2006}. However, during the last few years of
inspiral, when the small object is orbiting deep inside the SMBH's
gravitational field, GWs are radiated continuously in the LISA band.
During this phase, the residual eccentricity is typically at $e
\lesssim 0.4$~\cite{GHK,GHK2} and the orbital motion will exhibit
extreme versions of relativistic effects (e.g., periastron and
Lense-Thirring precession).  As a consequence, the resulting GW signal
will be strongly ``colored'' by these effects and will take a
complicated form~\cite{scott,zoom}.  By complicating the waveform,
these strong-field effects potentially make data analysis difficult;
but, they also encode a great deal of information about the strong
field nature of the spacetime. By accurately measuring all of these
effects, it is expected that we will be able to ``map'' the spacetime
of the black hole~\cite{ryan95}, probing its multipolar structure and
verifying that it obeys the constraints that general relativity
imposes on black hole solutions~\cite{bumpy,qkerr}.

The expected amplitude of the signal from these extreme-mass-ratio
inspirals (EMRIs) is about an order of magnitude below LISA's
projected instrumental noise and, at low frequencies ($\lsim 2$ mHz),
is several orders of magnitude below ``confusion noise'' produced by
unresolved Galactic binaries~\cite{Gbinaries}. However, the signals
will be observed for $\sim 10^5$ waveform cycles, and matched
filtering will therefore allow detection of these signals with high
signal-to-noise ratio (SNR) out to a redshift $z\sim1$~\cite{rates}.
Preliminary estimates suggest LISA could see as many as $\sim 10^3$
EMRI events during its lifetime \cite{rees2, rates}, using a suitable
``semi-coherent'' search algorithm and provided that confusion noise
can be efficiently reduced in the real data.

Matched filtering algorithms require the correlation of the detector's
output data stream with a bank of waveform templates which describe
the real signal with sufficient accuracy, covering the whole parameter
space. The fact that we are dealing with a binary system of extreme
mass ratio $\mu/M \ll 1$ means that the gravitational waveform may be
obtained accurately using black hole perturbation theory.  The extreme
mass ratio also guarantees that the orbital parameters evolve
adiabatically under radiation reaction; i.e., they evolve on a much
longer time scale than the orbital periods.  This implies that within
the radiation reaction time scale, the inspiral waveform can be
approximated by ``snapshot'' waveforms --- waveforms calculated by
assuming that the small object is moving along a geodesic, neglecting
backreaction for that short stretch of time.  These snapshots are
constructed using the Teukolsky equation {\cite{Teuk}}, an equation
that describes the first order change to the curvature tensor of a
black hole due to some perturbing source.  Accurate Teukolsky-based
(TB) snapshot waveforms have been calculated for inclined-circular
orbits \cite{scott}, equatorial-eccentric orbits
\cite{zoom,cutler,shibata} and most recently, for a certain number of
generic (inclined-eccentric) orbits~\cite{drasco}. The reader can find
recent reviews on the modeling of EMRI waveforms and orbital dynamics
in Refs.~\cite{kg_review,chapter}.

TB waveforms are computationally expensive to generate as they require
the numerical integration of the Teukolsky equation and summation over
a large number of multipole modes. In this sense, these waveforms are
not ``user-friendly'', especially when it comes to realistic data
analysis calculations where one has to handle a bank which contains
$\sim 10^{12}$ of these waveforms~\cite{rates}.  Moreover, the
Teukolsky formalism does not provide any information on
``conservative'' self-interaction effects.  To compute these, one must
use a self-force formalism.  This approach is still under development
and is very likely to be even more computationally expensive when it
is completed (see~\cite{poissonLR} for a recent review).

These difficulties have led to a need for the construction of
approximate families of waveforms that capture the main features of
the true signals, but which can be generated quickly in large
numbers. Such approximate waveforms are already being used for scoping
out data analysis algorithms for the detection of EMRIs in LISA data
\cite{rates}, and may ultimately play some role as fiducial detection
templates in the final data analysis.  One possible approach is to
construct post-Newtonian (PN) waveforms, which have the advantage of
being analytic and therefore very easy to generate. Post-Newtonian
EMRI waveforms have been computed in the Schwarzschild spacetime for
both quasi-circular \cite{poisson} and eccentric orbits~\cite{Levin}, and in
the spacetime outside a slowly rotating Kerr black hole for 
quasi-circular orbits~\cite{Pois2}.

Recently another class of approximate waveforms have become available,
based on various ``kludge'' approaches~\cite{webpage}.  The basic idea
of the kludges is to combine different prescriptions for the orbital
evolution and waveform emission (not necessarily in a self-consistent
way).  An ``analytic kludge'' (AK) was developed by Barack and Cutler
\cite{AK} (see also~\cite{garrido}).  In their model, the small object
is moving on a Keplerian orbit, amended to include the effects of
pericenter precession, Lense-Thirring precession, and inspiral from
radiation reaction. The emitted gravitational waveforms are described
by the lowest-order quadrupole formula.

In this paper, we consider an alternative way to construct kludge
waveforms.  This approach is much less amenable to analytic
calculation, so we called it the ``numerical kludge'' (NK).  The first
step of the NK is to produce an inspiral trajectory in ``phase
space''--- the space defined by the constants (orbital energy, axial
angular momentum, and ``Carter constant'') which specify orbits (up to
initial conditions)~\cite{GHK,GHK2}.  The second step is to
numerically integrate the Kerr geodesic equations along this inspiral
trajectory and hence obtain the Boyer-Lindquist coordinates of the
inspiralling object as a function of time~\cite{NK}.  The final step
is to construct a gravitational waveform from this inspiral
trajectory.

The approach to waveform construction that we take is to identify the
Boyer-Lindquist coordinates of the source with spherical polar
coordinates in flat-space. There are several different expressions
available in the literature for the gravitational waveforms from
orbits in flat space, and we apply these to our pseudo-flat-space
trajectory to construct waveforms. Specifically, we look at the
quadrupole formula~\cite{MTW}, which is valid in the limit of
weak-field (i.e., flat-space) and slow motion.  We also examine the
quadrupole-octupole formula of Bekenstein~\cite{bekenstein}, as well
as a formula derived by Press~\cite{press}.  The Press formula is also
a weak-field expression, but is not restricted to slow-motion or small
sources, and contains radiation at orders higher than quadrupole and
octupole.

The purpose of this paper to establish and delimit the accuracy and
reliability of waveforms constructed in these various ways.  We do
this by comparing to TB waveforms.  TB waveforms are currently the
most accurate EMRI waveforms available.  In most cases, TB waveforms
represent the emission from geodesic orbits --- we mostly do not
include the radiative evolution of orbital parameters in this work.
(With one exception: because complete TB inspirals exist for
zero-eccentricity orbits {\cite{scott_insp}}, we compare to a full
inspiral in this case.)  We compare the various NK waveforms with TB
waveforms using an overlap integral which weights the waveforms in
frequency space by the expected LISA noise curve, and maximizes the
overlap with respect to time offsets.  This overlap is identical to
the test used to evaluate the efficiency of model waveforms as
detection templates.

We find that quadrupole-octupole NK waveforms are in remarkable
agreement (overlaps$~\gtrsim 0.95$) with TB waveforms for orbits with
periastron $r_p \gtrsim 5M$. Most orbits in the final year of the
inspiral satisfy this restriction which means that NK waveforms are
quite accurate over a considerable portion of the inspiral parameter
space. For orbits that come below that radius (mostly prograde orbits
around rapidly spinning holes), the agreement rapidly degrades
(although NK waveforms remain more accurate than post-Newtonian or AK
waveforms). This is not surprising --- for such orbits the full TB
waveform receives significant contributions from higher multipoles and
back-scattering from the spacetime.  These effects are {\it ab initio}
absent in the NK prescription. Waveforms generated using the Press
formula perform similarly well for $r_p \gtrsim 5M$, and slightly
better than the quadrupole-octupole prescription for orbits where the
contribution from higher harmonics is not negligible.  The improvement
in this relatively strong-field regime is only slight, however --- we
did not find any regime where the Press formula is a significant
improvement on the quadrupole-octupole prescription.
We conclude that NK waveforms can accurately reproduce true
gravitational waveforms in a large part of parameter space; but, there
is little gain from going beyond the quadrupole-octupole prescription.

We also calculate kludge radiation {\it fluxes} by combining the
quadrupole energy flux formula with exact geodesic motion and
averaging in time; such an analysis was done for Schwarzschild orbits
in~\cite{GKL}. The resulting fluxes typically compare well to the best
available PN formula in the weak field, but provide a somewhat better
estimate of the energy and angular momentum fluxes (comparing to TB
results) for strong-field orbits.  These fluxes also allows us to
assess by how much the approximations which go into the kludge
construction are inconsistent.  Since the inspiral trajectory and
waveform construction are considered separately, kludge GWs carry a
different amount of energy and angular momentum to infinity than the
inspiralling particle loses (which is set by the formulas which
determine the inspiral through orbital parameter space). It is
important to bear this in mind when using kludge waveforms in
computations, e.g., for estimating SNRs of LISA EMRI detections, and
to have an estimate of the size of the inconsistency.  Finally, we
describe how the kludge waveforms may be improved by including some of
the conservative self-force effects, which may have a significant
imprint on true inspiral waveforms.

This paper is organised as follows. Section~\ref{inventory} provides a
review of existing EMRI waveforms, paying special attention to the
kludge semi-relativistic waveforms.  Details of how these waveforms
can be generated are described. In Section~\ref{overlaps} we discuss
the overlap function between two waveforms, as this will be our main
tool for comparison.  Section~\ref{results} contains the results from
the comparison between kludge and Teukolsky-based waveforms and
fluxes. In Section~\ref{cons_pieces} a method is outlined for
including the conservative parts of the self-force in our kludge
scheme. Finally, we present a concluding discussion in
Section~\ref{conclusions}. We shall use greek letters ($\mu$, $\nu$,
etc.) to denote spacetime indices, and latin letters ($i$, $j$, etc.)
to denote spatial indices (unless explicitly stated
otherwise). Throughout the paper we adopt geometric units $G=c=1$.


\section{Waveform inventory}
\label{inventory}

Presently, several types of EMRI waveforms are available. In broad
terms, these waveforms fall into three categories: (i) those
calculated numerically within the framework of black hole perturbation
theory (Teukolsky-based, or TB waveforms), (ii) analytic waveforms
which result from self-consistent PN expansions of the GW equations
{\it and} orbital motion, and (iii) approximate semi-relativistic
waveforms, or ``kludges''.  This third category is the focus of this
paper.

Kludge waveforms are constructed by combining a flat spacetime
wave-emission formula with either a fully relativistic treatment of
particle motion (resulting in the numerical kludge, or NK, waveforms),
or some analytic approximation of this motion (leading to the analytic
kludge, or AK, waveforms).  We shall examine the construction of NK
waveforms, comparing them to TB waveforms (the most accurate EMRI
waveforms presently available).  For completeness and background to
this paper's discussions, we now briefly discuss each of the above
waveform families.


\subsection{Teukolsky-based numerical waveforms}

The primary framework for black hole perturbation theory in a Kerr
background is the Teukolsky formalism~\cite{Teuk}, which encapsulates
all gravitational radiative degrees of freedom in a single ``master''
wave equation --- the ``Teukolsky equation'' --- for the Weyl scalars
$\psi_0$ and $\psi_4$. A key feature of this equation is that it
admits separation of variables in the frequency domain, which
effectively reduces it to a pair of ordinary differential
equations. There are extensive results in the literature on solutions
of the Teukolsky equation, starting from Press and
Teukolsky~\cite{PT}, Detweiler~\cite{detweiler} and Sasaki and
Nakamura~\cite{SN}.  More recent work uses the
Teukolsky-Sasaki-Nakamura formalism, see Refs.~\cite{kg_review,
chapter} for detailed discussions and references on the subject.

To date, the Teukolsky equation has been solved in the frequency
domain for circular-inclined orbits~\cite{scott}, eccentric-equatorial
orbits~\cite{zoom} and most recently for a handful of generic
(eccentric-inclined) orbits~\cite{drasco}.  We make use of the
waveforms generated by these various authors in this paper to assess
the quality of our NK waveforms. Recently, the Teukolsky equation has
also been solved directly in the time domain
\cite{martel,ramon}. Time-domain calculations have the great advantage
of speed, since they avoid the need for Fourier decomposition and
summation over all orbital frequency harmonics.  However, these
calculations are not yet mature enough to provide accurate waveforms
from Kerr orbits, because of the difficulty of representing the
various $\delta$-functions appearing in the energy-momentum tensor of
a point particle.

As a final remark, we should mention again that in all the above TB
calculations (either in the frequency or the time domain) the motion
of the small object is taken to be strictly geodesic. This is a
reasonable first approximation since for an EMRI system the orbital
evolution is adiabatic, i.e., over a time interval of several orbits
the motion is geodesic to high precision. However, for longer
stretches of time ($\sim M^2/\mu$) the effects of gravitational
back-reaction become significant and cannot be ignored anymore.
Waveforms that take into account an evolving orbit (and the
conservative self-interaction) require the computation of the
gravitational self-force (see~\cite{poisson_rev} for an up-to-date
review and a full list of references). However, self-force waveforms
are not yet available and are unlikely to be for the next few years.
Moreover, it is very likely that self-force calculations will remain
quite computationally intensive; as such, it is very likely that it
will not be possible to generate self-force based waveforms in
sufficient numbers to be used for LISA data analysis. It is therefore
essential to investigate approximate, easy-to-use waveform models.

\subsection{Analytic waveforms}

Most available analytic waveforms are based on post-Newtonian
expansions of the orbital dynamics and wave emission, an expansion
that is of greatest relevance when the bodies are widely separated.
These waveforms are typically constructed for a specific object or for
restricted orbital motion. A significant amount of work has been done
on modeling waveforms from two spinning bodies with comparable masses
orbiting in precessing quasi-circular orbits (a key GW source for both
ground- and space-based detectors).  Kidder~\cite{Kidder} has
investigated the effects of spin-spin and spin-orbital coupling on the
waveform from inspiralling binaries. The most recent investigation of
PN waveforms (and their application to data analysis) for spinning
binary systems may be found in Refs.~\cite{BCV2, PBCV, BlBuon} (see
also references therein).  Another promising approach is described
in~\cite{spinEOB}, which is the first attempt to extend the
``effective-one-body'' method~\cite{EOB} to spinning compact objects.
In parallel to modeling spinning binaries in quasi-circular orbits,
there has been significant progress in the construction of waveforms
for eccentric, comparable mass binaries~\cite{GopIyer1, 3.5eccen,
Mem}. Presently, there are no post-Newtonian waveforms which include
both eccentricity of the orbit and spins of the orbiting bodies.

Post-Newtonian models are ultimately not so useful for modeling EMRIs,
since most of the GWs observable to LISA are generated from a strong
field region ($r\sim $ a few $M$), where the PN expansion is unlikely
to be reliable.  One can, however, construct PN waveforms in the EMRI
limit, accepting their certain unreliability as a way to develop a
very ``quick and dirty'' set of tools for studying these waves.  PN
EMRI waveforms are available for systems of a test mass in a
quasi-circular~\cite{poisson} or eccentric~\cite{Levin} orbit around a
Schwarzschild black hole, or in a quasi-circular orbit around a slowly
rotating Kerr black hole~\cite{Pois2}.

More recently, a class of approximate PN waveforms has been developed
by Barack and Cutler {\cite{AK}}.  These ``analytic kludge'' (AK)
waveforms are essentially phenomenological waveforms --- they are
constructed using the classic quadrupole waveforms for eccentric
Keplerian orbits derived in~\cite{matthews}, but with the relativistic
effects of pericenter precession, Lense-Thirring precession, and
inspiral imposed.  Though not as accurate as the NK waveforms
described in this paper, the AK are very quick to generate, and have
proven to be useful for computing the Fisher information matrix in
investigations of parameter measurement with EMRI GWs {\cite{AK}}.

The overlap between AK and NK waveforms is high in the very weak field, but degrades as the orbital periapse is decreased (see \cite{rates}; more detailed comparisons will be included in future papers on the semi-coherent algorithm currently in preparation). Even for geodesic orbits, AK and NK waveforms with the {\it same physical parameters} drift out of phase quickly, since the frequency structure of the two waveform families is different. This arises because the AK uses a Keplerian orbital parameterisation, compared to the geodesic parameterisation used in the NK. For an equatorial orbit with semi-latus rectum $p=30M$ and eccentricity $e=0.3$ around a $10^6M_{\odot}$ black hole of spin $a=0.8$, the azimuthal frequency of the NK waveform is $0.196$mHz compared to $0.216$mHz for the AK. These orbits will therefore be one cycle out of phase within $\sim 6$ hours. AK waveforms will thus not be particularly {\it faithful} templates. The problems can be mitigated by adjusting the orbital parameters of the AK waveform to improve the match with the NK, and the AK waveforms do capture the main features of true EMRI waveforms. For this reason, they may be quite {\it effectual} templates, but this has not yet been properly assessed. In the future, the effectualness of the AK waveforms as detection templates will be investigated by using banks of AK templates to search for more accurate NK or TB waveforms embedded in noise.


\subsection{Semi-relativistic numerical ``kludge'' waveforms}

The idea behind the numerical kludge (NK) waveforms is to combine an
exact particle trajectory (up to inaccuracies in the phase space
trajectory and conservative radiation reaction terms) with an
approximate expression for the GW emission. By including the particle
dynamics accurately, we hope to capture the main features of the
waveform accurately, even if we are using an approximation for the
waveform construction.

The computation of NK waveforms proceeds in three steps.  The first is
to construct an inspiral trajectory in ``phase space'' --- that is,
the space of constants $E$ (energy), $L_z$ (axial angular momentum)
and $Q$ (Carter constant) which characterizes Kerr black hole orbits
(up to initial conditions).  The construction of the phase space
trajectory has already been described in previous work
\cite{GHK,GHK2}.  In this paper we shall (in most cases) ignore the
evolution of orbits due to radiation reactions.  The procedure for
waveform construction including orbital evolution is identical to that
given here.  The second step is to integrate the Kerr geodesic
equations along the inspiral trajectory and hence obtain the
Boyer-Lindquist coordinates of the inspiralling object as a function
of time~\cite{NK}. The final step is to construct a gravitational
waveform from this inspiral trajectory. We do this by identifying the
Boyer-Lindquist coordinates ($r, \theta, \phi, t$) with spherical
polar coordinates in flat-space and then evaluating a flat-space
emission formula for the corresponding flat-space source orbit.

The idea of coupling a weak field formula with fully relativistic
motion first appeared in papers by Sasaki and Ruffini~\cite{ruffini,
sasaki} and was then investigated more thoroughly for a test-body
orbiting a Schwarzschild black hole by Tanaka et al.\
\cite{shibata}. More recently, Gair et al.\ \cite{GKL} looked at
semi-relativistic fluxes for Schwarzschild orbits and derived analytic
expressions for the fluxes from arbitrary orbits. Their focus was on
highly eccentric orbits relevant to the capture problem. In all these
cases, the authors were interested in computing semi-relativistic
fluxes, rather than waveforms. While our focus is on waveforms rather
than fluxes, in Section~\ref{results_flux} we do calculate
semi-relativistic fluxes for Kerr orbits, as these results have not
yet appeared in the literature.

In the remainder of this section, we describe the two stages of
waveform generation (ignoring backreaction): (i) the computation of a
trajectory, and (ii) the computation of a gravitational waveform from
an arbitrary trajectory.

\subsubsection{Computing the orbital trajectory}
\label{insptraj} The first step in constructing a numerical kludge
waveform is to compute the trajectory that the inspiralling body
follows in the Boyer-Lindquist coordinates of the Kerr spacetime of
the central black hole. Ignoring radiation reaction, this path is a
Kerr geodesic. Geodesic motion in the Kerr space-time is
well-known~\cite{bardeen,chandra} and is governed by a set of
first-order differential equations:
\begin{subeqnarray}
\Sigma \frac{dr}{d\tau} &=& \pm \sqrt{V_r}, \\
\Sigma \frac{d\theta}{d\tau} &=& \pm \sqrt{V_{\theta}}, \\
\Sigma \frac{d\phi}{d\tau} &=&  V_\phi,\slabel{geoEqphi} \\
\Sigma \frac{dt}{d\tau} &=&  V_t, \slabel{geoEqt}
\end{subeqnarray}
where the various  ``potentials'' are defined by
\begin{subeqnarray}
V_r &=& \left[ E(r^2 + a^2) - L_z a \right]^2 -
\Delta\left[ r^2 + (L_z-aE)^2 + Q\right],\\
V_{\theta} &=&  Q -\cos^2{\theta}\left[  a^2(1-E^2)
+ \frac{L_z^2}{\sin^2{\theta}} \right],\\
V_{\phi} &=& \frac{L_z}{\sin^2{\theta}} - aE
+ \frac{a}{\Delta}\left[  E(r^2+a^2) -L_za \right],\\
V_t &=& a \left(L_z - aE\sin^2{\theta}\right)
+ \frac{r^2+a^2}{\Delta} \left[ E(r^2+a^2) - L_za \right].
\end{subeqnarray}
Here, $ \Sigma = r^2 + a^2\cos^2{\theta}$, and $\Delta = r^2 -2Mr +
a^2$. The constants $E$, $L_z$, $Q$ are the three first integrals of
the motion: $E$ is the orbital energy; $L_z$ is the projection of the
orbital angular momentum along the black hole's spin axis; and $Q$ is
known as the ``Carter constant''.  This last constant is the
relativistic generalization of the ``third integral'' used to separate
the equations which describe orbits in an axisymmetric gravitational
potential (a result which is particularly well-known in the literature
describing orbits in galactic potentials~\cite{binneytremaine}).  In
the spherical (i.e., Schwarzschild) limit, $Q$ reduces to the square
of the orbital angular momentum projected into the equatorial plane;
see~\cite{GHK} and references therein for discussion.

For a given $E$, $L_z$ and $Q$, the roots of $V_r$ determine the
turning points of the radial motion---the periastron $r_p$, and
apastron $r_a$. From these, one can define an orbital eccentricity
$e$, and semi-latus rectum $p$, using the conventional Keplerian
definitions
\begin{subeqnarray}
\label{pedef}
r_{p}=\frac{p}{1+e}, &\quad&  r_{a}=\frac{p}{1-e},\\
\Rightarrow \quad
p = \frac{2\,r_{a}\,r_{p}}{r_{a}+r_{p}}, &\quad&
e=\frac{r_{a}-r_{p}}{r_{a}+r_{p}}.
\end{subeqnarray}
We also replace the Carter constant by an ``inclination angle'',
defined by
\be
\tan^2\iota = \frac{Q}{L_z^2}. \label{iotadef}
\en
To aid numerical integration, one can work in terms of two angular
variables, $\psi$ and $\chi$, instead of $r$ and $\theta$. We define
$\psi$ by the equation
\begin{equation}
r=\frac{p}{1+e\cos\psi}. \label{rofpsi}
\end{equation}
We define $\chi$ by the equation $z=\cos^2\theta=z_{-}\cos^{2}\chi$,
where $z_{-}$ is given by
\begin{equation}
\beta(z_{+}-z)(z_{-}-z)
=   \beta z^{2}-z\left[Q+L_{z}^{2}+a^{2}(1-E^{2})\right] +Q,
\end{equation}
with $\beta=a^{2}(1-E^{2})$. Expanding the radial potential as
\begin{equation}
V_r=(1-E^{2})\,(r_{a}-r)\,(r-r_{p})\,(r-r_{3})\,(r-r_{4}) ,
\end{equation}
we find evolution equations for $\psi$ and $\chi$ of the form
\begin{eqnarray}
\frac{\rmd \psi}{\rmd t} &=& \frac{\sqrt{1 - E^{2}}\,\left[(p -
r_{3}(1 - e)) - e(p + r_{3}(1 -
e)\cos\psi)\right]^{\frac{1}{2}}\,\left[(p - r_{4}(1 + e)) + e(p -
r_{4}(1 + e)\cos\psi)\right]^{\frac{1}{2}}}{\left[\gamma +
a^{2}\,E\,z(\chi)\right](1 - e^{2})},
\label{dpsidt} \\
\frac{\rmd \chi}{\rmd t} &=& \frac{\sqrt{\beta\,\left[z_{+} -
z(\chi)\right]}}{\gamma + a^{2}\,E\,z(\chi)}
\label{dchidt} \\
\nonumber {\rm where} \hspace{0.1in}\gamma &=& E\left[\frac{\left(r^2
 + a^2\right)^{2}}{\Delta} - a^2 \right]
 -\frac{2\,M\,r\,a\,L_{z}}{\Delta} .
\end{eqnarray}
In terms of the variables $\phi$, $\psi$ and $\chi$, the geodesic
equations are well behaved at the turning points of the motion, which
facilitates numerical integration (as discussed in~\cite{NK}).

Although we mostly ignore it here, it is easy to include radiation
reaction in this prescription. To do so, one first computes an
inspiral trajectory through phase space by writing
\begin{subeqnarray}
\label{inspeqns}
\frac{\rmd E}{\rmd t} &=& f_E (a, M, m, p, e, \iota), \\
\frac{\rmd L_z}{\rmd t} &=& f_L (a, M, m, p, e, \iota), \\
\frac{\rmd Q}{\rmd t}&=& f_Q (a, M, m, p, e, \iota).
\end{subeqnarray}
The functions $f_E$, $f_L$ and $f_Q$ are derived from well-known
post-Newtonian results, augmented by additional corrections. The
leading order part of $f_E$ is given in a later section of this paper,
Eq.~\erf{GHKEdot}.  (For the full expressions see
Ref.~\cite{GHK2}). Since the parameters $a$, $M$ and $m$ are constant,
and $p$, $e$ and $\iota$ are directly related to $E$, $L_z$ and $Q$
via Eqs.~\erf{pedef}-\erf{iotadef}, equations~\erf{inspeqns} can be
integrated to give the phase space trajectory. That is, the values of
$E$, $L_z$ and $Q$ (or equivalently $p$, $e$ and $\iota$) are given as
functions of time.  To obtain the trajectory of the inspiralling
particle, these time dependent expressions can be substituted into the
right hand sides of Eqs.~\erf{geoEqphi}, \erf{geoEqt}, \erf{dpsidt}
and~\erf{dchidt}.  The resulting set of ODEs can then be integrated to
give the inspiral trajectory. In the following (with one exception),
we will only consider waveforms from geodesic trajectories (setting
$f_E \equiv f_L \equiv f_Q \equiv 0$), and refer the reader to
Refs.~\cite{GHK,GHK2} for more details on the construction of
inspirals.

Once a trajectory has been obtained in this manner, one constructs
an ``equivalent'' flat-space trajectory by projecting the
Boyer-Lindquist coordinates onto a fictitious spherical polar
coordinate grid, defining the corresponding Cartesian coordinate
system and pretending that these new coordinates are true flat-space
Cartesian coordinates:
\begin{subeqnarray}
x &=& r \sin\theta \cos\phi,\\
y &=& r \sin\theta \sin\phi,\\
z &=& r \cos\theta.
\end{subeqnarray}
We use the resulting flat space trajectory as input to a wave
generation formula. This is a ``bead on a wire'' prescription---by
putting the trajectory in flat space, we remove the background that is
causing the curvature of the geodesic path and hence we are forcing
the particle to move along a curved path like a bead moving on a
wire. This leads to obvious inconsistencies in the approach --- e.g.,
the non-conservation of the flat-space energy-momentum tensor of the
particle motion since we are not including the energy-momentum of
``the wire'' along which the particle moves.

\subsubsection{Waveform generation}

Having constructed the particle orbit in our pseudo-flat-space, we now
apply a flat-space wave generation formula. Consider the weak-field
situation and write down the spacetime metric as $g_{\mu\nu} =
\eta_{\mu\nu} + h_{\mu\nu}$, where $\eta_{\mu\nu}$ is the flat metric
and $h_{\mu\nu}$ are small perturbations.  The {\it trace-reversed}
metric perturbation is defined as $\bar h^{\mu\nu}\equiv h^{\mu\nu} -
(1/2)\eta^{\mu\nu} h$, where $h = \eta^{\mu\nu}h_{\mu\nu}$. Imposing
the Lorentz gauge condition $\bar{h}^{\mu\alpha}_{\ \ ,\alpha}=0$, the
linearized Einstein field equations can be written as
\begin{eqnarray}
\Box \bar{h}^{\mu \nu}&=& -16\pi {\cal T}^{\mu \nu} \label{waveq},
\end{eqnarray}
in which $\Box$ denotes the usual flat space wave operator and the
effective energy-momentum tensor ${\cal T}^{\mu\nu}$ satisfies
\begin{eqnarray}
{\cal T}^{\mu\nu}\mbox{}_{,\nu} &=& 0. \label{emcons}
\end{eqnarray}
Here a comma subscript denotes partial derivative ($f_{,\mu} =
\partial f/\partial x^\mu$). Note that our source conservation
equation uses a partial rather than a covariant derivative.  This is
because we would hope, in a self-consistent approach, to choose
coordinates so that the energy momentum tensor is flat-space
conserved.  Finally, when observing GWs at large distances, we are
really only interested in the transverse and traceless parts of the
spatial components of $\bar{h}^{jk}$; a projection of these components
will be necessary.

Taking coordinates centered at the black hole, we denote the position
of the observer by $(t,{\bf x})$ and the position of the particle by
$(t_p,{\bf x}_p)$).  The wave equation~\erf{waveq} has the familiar
retarded time solution
\begin{equation}
\bar{h}^{jk}(t,{\bf x})
=   4 \int \frac{{\cal T}^{jk}(t-|{\bf x} - {\bf x'}|, {\bf x'})}
    {|{\bf x} - {\bf x'}|} \;\; \rmd^{3}x'.
\label{retint}
\end{equation}
The additional coordinate ${\bf x'}$ is the integration variable,
which goes over every possible space location where the effective
energy-momentum tensor ${\cal T}^{\mu\nu}(t', \bx')$ is nonzero. If
the source motion is only negligibly influenced by gravity, then
${\cal T}^{\mu\nu}$ may be taken to equal $T^{\mu\nu}$, the
energy-momentum tensor of the matter source. In Ref.~\cite{press},
Press derived a formula valid for extended, fast motion sources.  This
was obtained by substituting Eq.~\erf{emcons} into Eq.~\erf{retint}
repeatedly. The result is
\begin{equation}
\bar{h}^{jk} (t,\bx)
=   \frac{2}{r} \frac{\rmd^{2}}{\rmd t^{2}}
    \int \left[ \left({\cal T}^{00} - 2{\cal T}^{0l}n_{l} +
    {\cal T}^{lm}n_{l}n_{m} \right) x'^{j}x'^{k} \right]_{t'=t-|\bx-\bxd|}
    \rmd^{3} x',
    \label{press}
\end{equation}
in which $r^{2} = \bx\cdot \bx$ and ${\bf n} = \bx /r$. In the slow
motion limit, the Press formula reduces to the usual quadrupole
formula (hereafter, an overdot denotes a time-derivative)
\begin{eqnarray}
\bar{h}^{jk}(t, \bx)
=   \frac2{r} \left[ \ddot{I}^{jk}(t')  \right]_{t'=t-r},\
    \label{quad}
\end{eqnarray}
where
\begin{eqnarray}
I^{jk}(t') = \int x'^{j}x'^{k} T^{00}(t',{\bf x'}) d^3x'
\label{quadmoment}
\end{eqnarray}
is the source's mass quadrupole moment. Including the next order terms
(the mass octupole and current quadrupole moments of the source,
denoted $M^{ijk}$ and $S^{ijk}$ respectively), we obtain the
quadrupole-octupole formula (see Refs.~\cite{bekenstein, press} for
details),
\begin{eqnarray}
\bar{h}^{jk} =
    \frac2{r} \left[ \ddot{I}^{jk} - 2 n_i \ddot{S}^{ijk} +
    n_i \dddot{M}^{ijk}\right]_{t'=t-r},\ \label{octup}
\end{eqnarray}
with
\begin{eqnarray}
S^{ijk}(t')
&=& \int x'^{j}x'^{k} T^{0i}(t',{\bf x'}) d^3x' ,
    \label{currquaddef} \\
M^{ijk}(t')
&=& \int x'^i x'^j x'^k T^{00}(t',{\bf x'}) d^3x'.
    \label{massoctdef}
\end{eqnarray}
Note that in both Eqs.~(\ref{quad}) and (\ref{octup}), the retarded
time is $t-r$ instead of the more complicated expression appearing
in~(\ref{press}). If desired, it is a straightforward (but
increasingly tedious) task to include more terms in the slow-motion
expansion of~(\ref{press}). In the present work we shall make use of
Eqs.~(\ref{quad}), (\ref{octup}) and the full Press
formula~\erf{press}.

The waveform in the standard transverse-traceless (TT) gauge is
given by the TT projection of the above expressions. We define an orthonormal 
spherical coordinate system via
\be
{\bf e}_r = \frac{\partial}{\partial r}, \quad {\bf e}_{\Theta} = \frac{1}{r}
\frac{\partial}{\partial\Theta}, \quad {\bf e}_{\Phi} = \frac{1}{r\sin\Theta}
\frac{\partial}{\partial\Phi}.
\en
The angles $\{ \Theta, \Phi \}$ denote the observation point's
latitude and azimuth respectively. The waveform in transverse-traceless gauge 
is then given by
\begin{equation}
h^{jk}_{TT} = \frac1{2}\left(
       \begin{array}{ccc}
       0 & 0 & 0\\
       0 & h^{\Theta\Theta}-h^{\Phi\Phi}& 2h^{\Theta\Phi}\\
       0 & 2h^{\Theta\Phi} & h^{\Phi\Phi}-h^{\Theta\Theta}
        \end{array}
  \right),
\end{equation}
with
\begin{subeqnarray}
h^{\Theta\Theta}
&=& \cos^2{\Theta} \left[ h^{xx} \cos^2{\Phi} + h^{xy}\sin{2\Phi} +
    h^{yy}\sin^2{\Phi} \right] + h^{zz}\sin^2{\Theta} -
    \sin{2\Theta}\left[ h^{xz}\cos{\Phi} + h^{yz}\sin{\Phi}\right],\\
h^{\Phi\Theta}
&=& \cos{\Theta} \left[ -\frac1{2} h^{xx}\sin{2\Phi} +
    h^{xy}\cos{2\Phi} +\frac1{2}h^{yy}\sin{2\Phi} \right] +
    \sin{\Theta}\left[ h^{xz}\sin{\Phi} - h^{yz}\cos{\Phi} \right],\\
h^{\Phi\Phi}
&=& h^{xx}\sin^2{\Phi} - h^{xy}\sin{2\Phi} + h^{yy}\cos^2{\Phi}.
\end{subeqnarray}
The usual ``plus'' and ``cross'' waveform polarizations are given by 
$h^{\Theta\Theta}-h^{\Phi\Phi}$ and $2h^{\Theta\Phi}$ respectively.
The expressions~\erf{press}, \erf{quad} and~\erf{octup} are valid for
a general extended source in flat-space. If we specialize to the case
of a point-mass $\mu$ moving along a trajectory $x'_p(\tau)$, we know
that the energy-momentum tensor in flat spacetime is given by
\begin{equation}
T^{\mu\nu} (t',{\bf x'})
=   \mu \int_{-\infty}^{\infty} \frac{\rmd x'^{\mu}_p}{\rmd \tau}
    \frac{\rmd x'^{\nu}_p}{\rmd \tau} \delta^{4}
    \left(x'-x'_{p}(\tau)\right) \;\; \rmd \tau
 =
    \mu \frac{\rmd \tau}{\rmd t'_p} \frac{\rmd x'^{\mu}_p}{\rmd \tau}
    \frac{\rmd x'^{\nu}_p}{\rmd \tau}
    \delta^{3} \left( \bxd - \bxd_{p}(t') \right),
    \label{emt}
\end{equation}
where $\tau$ is the proper time along the trajectory. It is related to
the particle's coordinate time by $\rmd\tau = (1-v^2/c^2)^{1/2}\, \rmd
t'$, where $v^2=|\rmd \bx'_p/\rmd t'_p|^2$. On the right hand side of
Eq.~\erf{emt} is a term $\rmd t'_p/\rmd \tau = 1 + O(v^2/c^2)$.
Expressions~\erf{quad} and~\erf{octup} are slow-motion expansions and
at the order of these expansions we should replace $\rmd t'_p/\rmd
\tau$ by $1$ for consistency.  The Press formula~\erf{press} is a
fast-motion expression, and therefore we do include this term. The
presence of a $\delta$-function in $T^{\mu\nu}$ facilitates the
simplification of the various moments in Eqs.~\erf{quad}
and~\erf{octup}:
\begin{eqnarray}
I^{jk} &=& \mu  x'^j_p x'^{k}_p\;,
\\
S^{ijk} &=& v^i\ I^{jk}\;,
\\
M^{ijk} &=& x'^i_p\ I^{jk}\;.
\end{eqnarray}
Here $v^a\equiv\rmd x'^a/\rmd t'_p$. The Press formula similarly
simplifies to
\begin{equation}
h^{jk} (t,\bx)
= \frac{2\mu}{r} \frac{\rmd^{2}}{\rmd t^{2}}
    \left[ \left(1 - n_a v^a \right) {x'}^{j}_p {x'}^{k}_p
           \left( \frac{\rmd t'_p}{\rmd \tau} \right)
    \right]_{t'_p =  t- |\bx -\bx'_p|} .
    \label{pressexp}
\end{equation}
The square bracketed expression is to be evaluated at a time $t'_p$
given implicitly by $t=t'_p+|\bx-\bxd_{p}(t'_p)|$. In a numerical
implementation, we evaluate the expression as a function of the time
$t'_p$ along the particle's path. In order to obtain a time series
that is evenly spaced in $t$ and compute the RHS of Eq.~\erf{pressexp}
by finite-differencing, we adjust the spacing of the sampling at the
particle, $\delta t'_p$, such that
\begin{equation}
\delta t = \left( 1 - n_a v^a \right) \delta t'_p. \label{dtrel}
\end{equation}
With a delta function source~\erf{emt}, the retarded-time
solution~\erf{retint} can be evaluated directly~\cite{ruffini}:
\begin{equation}
\bar{h}^{jk} (t,\bx)
= \left[\frac{4\mu}{r} \left(\frac{\rmd t'_p}{\rmd \tau} \right)
    \frac{\rmd {x'}^{j}_p}{\rmd t'_p}
    \frac{\rmd {x'}^{k}_p}{\rmd t'_p}
    \frac{1}{1-n_a v^a} \right]_{t'_p = t-|\bx-\bx'_p|}.
    \label{retexp}
\end{equation}
Naively, one could suppose that \erf{retexp} will perform better
than~\erf{press}, since it is derived using only one of the two
(invalid) flat space equations, rather than both.  In fact, we find
that the retarded integral expression~\erf{retexp} performs much worse
than either~\erf{quad}, \erf{octup} or~\erf{press} when compared to TB
waveforms.  The reason appears to be that the manipulations which lead
to the quadrupole, quadrupole-octupole, and Press formulae ensure that
the actual source terms --- mass motions --- are on the right hand
side.  This is to be contrasted with the retarded integral expression
\erf{retint}, which identifies the dominant GW on the left hand side
with weak spatial stresses on the right.

We must emphasize that the NK prescription is clearly inconsistent ---
we are binding the particle motion to a Kerr geodesic while assuming
flat spacetime for GW generation and propagation.  This is manifested
by the fact that the energy-momentum tensor~\erf{emt} is not flat
space conserved, $\partial_\nu T^{\mu\nu} \neq 0$.  However, the
spirit of this calculation is {\it not} a formal and consistent
approximation to EMRI waveforms; it is rather a ``phenomenological''
approach which takes into account those pieces of physics we believe
are the most crucial --- in particular, the exact Kerr geodesic
motion. By including the exact source trajectory, we ensure that the
spectral components of the kludge waveforms are at the correct
frequencies, although their relative amplitudes will be inaccurate. As
we shall see, this line of thinking is validated {\it post facto} by
the remarkable agreement between the kludge and TB waveforms (see
Section~\ref{results}).

It is important to underline the physical assumptions that have been
made in the derivation of (\ref{press}), (\ref{quad}) and
(\ref{octup}), in order to understand their generic limitations.
First of all, the assumed absence of any background gravitational
field means that our kludge waveforms are unable to capture any
features related to back-scattering. This effect is known to first
appear at 1.5PN [i.e. ${\cal O}(v^3)$] level (see,
e.g.,~\cite{damour}). Such ``tails'' of waves are particularly
prominent in the strong-field TB equatorial ``zoom-whirl''
waveforms~\cite{zoom} and in the waveforms from plunging or parabolic
orbits~\cite{kojima}. In all these cases, the hole's quasinormal mode
ringing adorns the emitted signal.

The slow-motion nature of expressions~(\ref{quad}) and~(\ref{octup})
suggests that they might be bad models for waveforms generated by
orbits venturing deep inside the central BH's spacetime.  The rich
multipole structure of the true waveform from such orbits is poorly
reproduced, as any slow-motion approximation essentially truncates the
sum over multipoles. The Press formula~(\ref{press}) includes
contributions at all multipoles and so might be expected to handle
these contributions quite well.  However, it turns out that it does
not perform that much better than the quadrupole-octupole waveform in
this regime. While it includes contributions at all multipoles, the
lack of background curvature in the waveform model means that the
amplitude of the higher modes is much lower than expected for true
EMRI waveforms.

Fortunately, these two deficiencies become important for the same
class of orbits---those that allow the body to approach very close to
the black hole. As a rule of the thumb we shall find that NK waveforms
are reliable (in terms of the overlap discussed in the next section)
as long as the closest orbital approach (periastron) is $r_p \gtrsim
5M $.



\section{The overlap between waveforms}
\label{overlaps}

We now give a brief description of the measure used for the
quantitative comparison between NK and TB waveforms. The main
motivation for the computation of accurate EMRI waveforms is to carry
out matched filtering for detection of the GWs. For the purpose of
signal detection and parameter estimation, a bank of templates is
constructed which covers the desired parameter space with sufficient
resolution. The detector output is then filtered through each
template. The measured strain amplitude in the detector, $x(t) = s(t)
+ n(t)$, consists of a (possibly present) signal $s(t)$, and the
detector noise $n(t)$. We define the Fourier representation of these
time series as
\begin{eqnarray}
\tilde{x}(f) = \int^\infty_{-\infty} x(t)e^{-i2\pi tf}dt.
\end{eqnarray}
The signal-to-noise ratio (SNR) can be expressed in terms of an
inner product defined on the vector space of possible signals. Given
two vectors (time series), $x(t_k), h(t_k)$, we define the overlap
$(x|h)$ by the equation \cite{CutlerFlan, Bala}
\begin{eqnarray}
(x|h) =  2 \int_0^{+\infty} \frac{\tilde{x}(f)\tilde{h}^*(f) + \tilde{x}^*(f)
\tilde{h}(f)}{S_h(f)} df,
\label{inprod}
\end{eqnarray}
where $S_h(f)$ is the one-sided noise power spectral density (PSD) (in
this case, the noise PSD for the LISA detector) and an asterisk
denotes complex conjugation.  Considering $h(t_k, \lambda^\alpha)$ as
a template with parameters $\lambda^\alpha$ we can approximate the SNR
by \cite{CutlerFlan}
\begin{equation}
\frac{S}{N}[h(\lambda^{\alpha})] =
\frac{\langle(s|h)\rangle_n}{\sqrt{\langle(n|h)^2\rangle_n}} =
\frac{(s|h)}{\sqrt{(h|h)}}\;.
\end{equation}
The notation $\langle f\rangle_n$ means to ensemble average the
function $f$ over all possible noise realizations $n$.  The PSD in
the denominator of Eq.~(\ref{inprod}) serves to suppress those
frequency components of the signal at which the detector noise is
large.

The main tool which will be used in this paper is not the SNR, but
rather the overlap function $\mathcal{O}$. The overlap is defined as
an inner product between two normalized vectors/signals:
\begin{eqnarray}
\mathcal{O} = (\hat{s}|\hat{h})\label{olap}\;.
\end{eqnarray}
The normalization is chosen so that $(\hat{s}|\hat{s}) =
(\hat{h}|\hat{h}) =1$.  The overlap can be regarded as the inner
product between two unit vectors; it varies within $[-1,1]$. The
overlap is equal to $1$ if the two waveforms are identical, and it
equals zero if the two waveforms are orthogonal (for example, cosine
and sine signals).  The overlap is an appropriate measure of
``goodness-of-fit'' since we are interested in knowing how well the NK
approximate the behavior of TB EMRI waveforms.
In this context, it is important to include the noise properties of
the detector since it is no problem if a template has poor performance
at frequencies where detector noise is large.

For a fair comparison, we should choose the signal (TB waveform) and
template to have the same parameters\footnote{Note that here we study
not the fitting factors but the faithfulness of the NK waveforms as
compared to TB-based ones, i.e., how well an approximate waveform with
given parameters reproduces the ``true'' waveform with the {\it same}
physical parameters. A faithful bank of waveforms could be used for
parameter estimation, while for detection all that is required is an
``effectual'' template bank, i.e., one in which every true waveform is
well represented by one template, even if that template has very
different parameters~\cite{damour98}.}.  In practice, the start times
of our signals and templates were slightly mismatched, so we allow for
maximisation over starting time (``time of arrival'').  This
maximisation is commonly done in GW data analysis, and can be
accomplished very efficiently in the Fourier transform, since the time
offset corresponds to a phase shift in the frequency domain.
Accordingly, we modify the inner product $(x|h)$ slightly:
\begin{equation}
(x|h)
=   {\rm max}_{t}\left( \int \frac{\tilde{x}(f)\tilde{h}^*(f) +
    \tilde{x}^*(f) \tilde{h}(f)}{S_h(f)} e^{i2\pi ft} df \right)
    \label{max_t}
\end{equation}
where $t$ corresponds to the time difference.  We find that the maximum usually occurs with an time offset close to zero (typically a few bins).

\begin{figure}[ht]
\includegraphics[height=0.4\textheight, keepaspectratio=true]{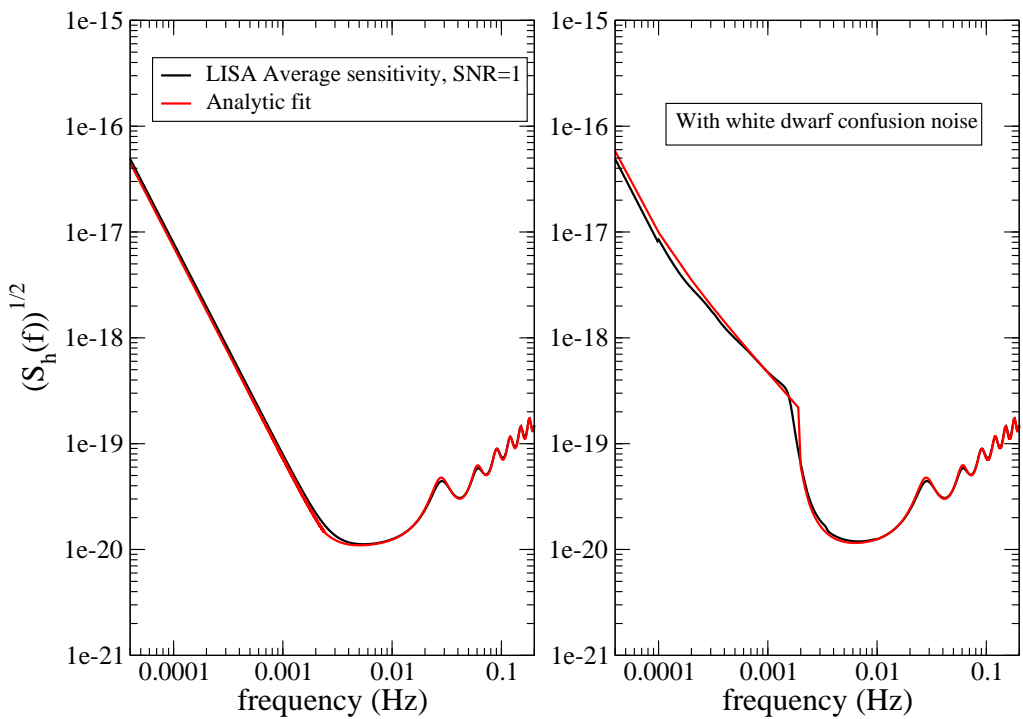}
\caption{Expected sensitivity curve $\sqrt{S_h(f)}$ for LISA; black curve:
numerical curve as generated in \cite{larson}, red curve: analytic
approximation used in this paper, see text for details).}
\label{S_h(f)}
\end{figure}

We conclude our discussion of the overlap by describing our
approximation to the LISA noise $S_h(f)$.  We have used an analytic
approximation to the numerically generated sensitivity curve given in
Ref.\ \cite{larson}. The agreement between these two curves is
excellent as clearly illustrated in Fig.~\ref{S_h(f)}. Our approximate
$S_h(f)$ function can be easily calculated according to the following
simple prescription. Define $u = 2\pi f\tau$, where $f$ is frequency
and $\tau = (5\times 10^6\,{\rm km})/c = 50/3$ sec is the light travel
time down one of LISA's arms.  For $u < u_{\rm trans} = 0.25$ we set
\begin{eqnarray}
S_h(f) = \frac{8.08\times 
10^{-48}}{(2\pi f)^4} +5.52\times10^{-41}.
\label{S1}
\end{eqnarray}
while for $u\ge u_{\rm trans}$,
\begin{eqnarray}
S_h(f) = \frac{1}{R}\left( \frac{2.88\times10^{-48}}{(2\pi f)^4}
+ 5.52\times10^{-41} \right),
\label{S2}
\end{eqnarray}
where
\begin{eqnarray}
R = \frac1{u^2}\left[ (1 + \cos^2(u))\left(\frac1{3} -
\frac2{u^2}\right) + \sin^2(u) + \frac{4\sin(u) \cos(u)}{u^3}
\right]\;.
\label{R}
\end{eqnarray}
Note that LISA's characteristics are incorporated in the light
travel time $\tau$ and in the numerical constants in the above 
expressions, so a similar mission but with different noise
characteristics would still be described by the functional form
defined by eqns.~(\ref{S1}-\ref{R}).

We have also included noise from a confused population of galactic
 white dwarf binaries following the prescription outlined in \cite{AK}.

\section{Kludge waveforms and fluxes: results}
\label{results}

\subsection{Comparison to Teukolsky-based geodesic waveforms}
\label{results_wav}
\subsubsection{Time-domain comparison}

In this Section we compare NK waveforms (using the overlap function
defined in the previous Section) to a variety of TB waveforms from
inclined-circular, equatorial-eccentric and some generic Kerr orbits.
The TB waveforms are treated here as ``exact'' (though this is not
strictly the case as discussed in Section~\ref{inventory}). Our
comparison rule is that the waveforms are generated using the same
Kerr orbit with identical initial conditions. We label these orbits by
the triad of orbital elements $\{p,e,\iota \}$ defined in equations
\erf{pedef}--\erf{iotadef}.

\begin{figure}
\includegraphics[height=0.33\textheight,keepaspectratio=true]{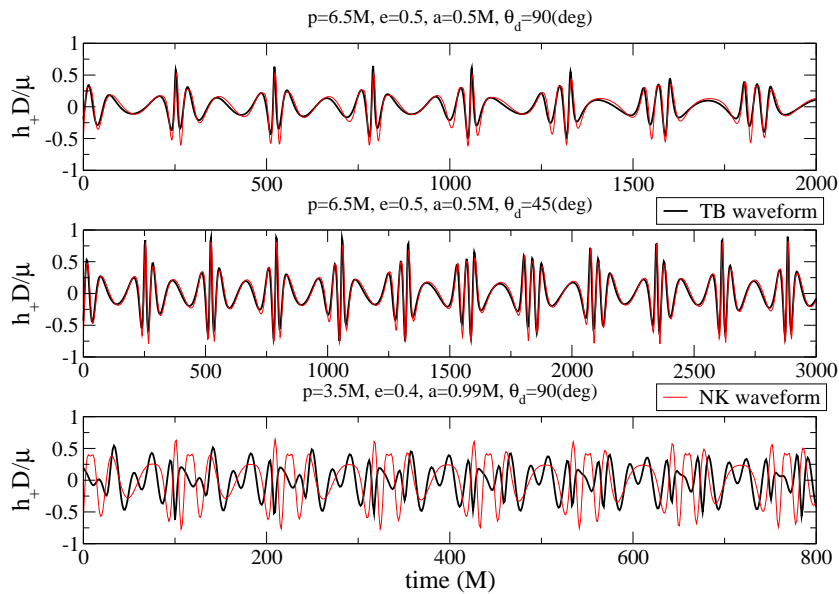}
\caption{Comparing TB and quadrupole-octupole kludge waveforms (black and red curves
 (bold black  and grey curves in the b\&w version),
respectively) for equatorial orbits and for an observer at a
latitudinal position $\theta = 45^\circ$ or $90^\circ$.  Orbital
parameters are listed above each graph.  The waveforms are
scaled in units of $D/\mu$ where $D$ is the radial distance of the
observation point from the source and $\mu$ is the test-body's
mass. The x-axis measures retarded time (in units of $M$) and we are
showing the `+' polarization of the GW in each case. The overlaps
between the NK and TB waveforms are $0.970, 0.987, 0.524$ going from
the top figure down.}
\label{equat4}
\end{figure}

\begin{figure}
\includegraphics[height=0.33\textheight,keepaspectratio=true]{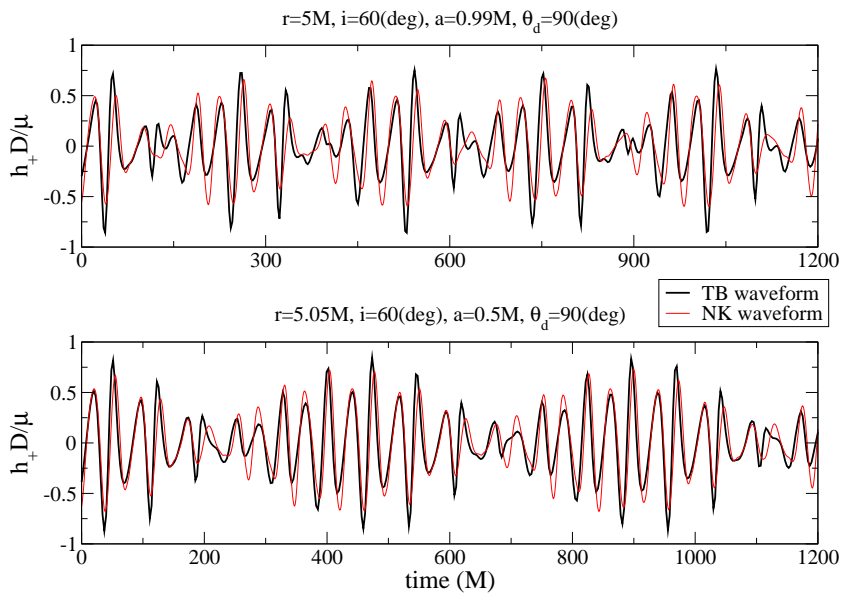}
\caption{Comparing TB and quadrupole-octupole kludge waveforms (black and red curves
 (bold black and grey curves in the b\&w version),
respectively) for circular-inclined orbits and for an observer at a
latitudinal position $\theta = 90^\circ$.  Orbital parameters are
listed above each graph.  The waveforms are scaled in units of
$D/\mu$ where $D$ is the radial distance of the observation point from
the source and $\mu$ is the test-body's mass. The x-axis measures
retarded time (in units of $M$) and we again show the plus polarization of the GW. 
The overlaps between the NK and TB waveforms are $0.882$ for
the top figure and $0.955$ for the bottom figure.}
\label{circ3}
\end{figure}

\begin{figure}
\includegraphics[height=0.34\textheight,keepaspectratio=true]{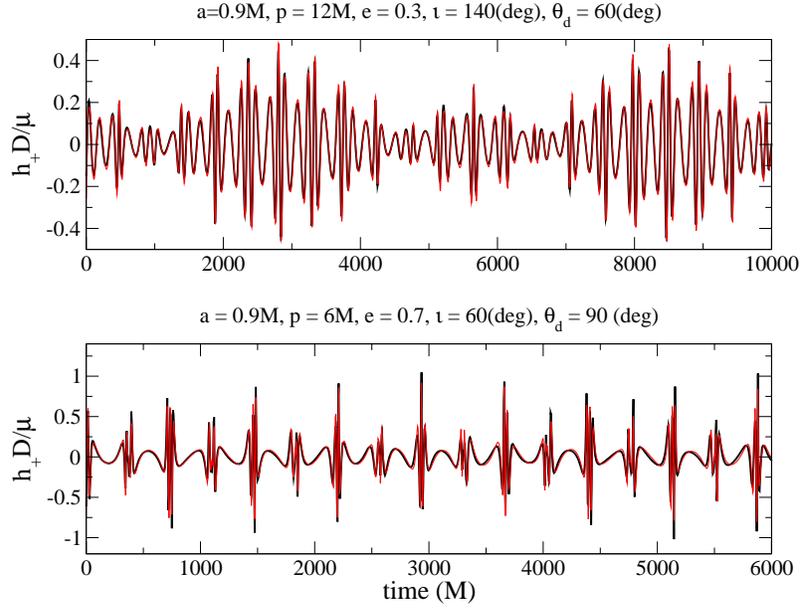}
\caption{Comparing TB and quadrupole-octupole kludge waveforms (black and red curves
 (bold black and grey curves in the b\&w version),
respectively) for generic orbits.  Orbital parameters are listed above
each graph.  The waveforms are scaled in units of $D/\mu$ where $D$ is the
radial distance of the observation point from the source and $\mu$ is
the test-body's mass. The x-axis measures retarded time (in units of
$M$). The overlaps between the NK and TB waveforms are $0.991$ and
$0.966$ for the top and bottom figures respectively.}
\label{gener}
\end{figure}

Figures \ref{equat4}, \ref{circ3}, and \ref{gener} show time-domain
waveforms for a selection of orbits in the moderate and strong
field. The NK waveforms plotted in these figures were generated by
using the quadrupole-octupole formula \erf{octup}. The results for
``plus'' and ``cross'' polarizations are essentially the same, so in
the figures we show only the ``plus'' polarization.  For all of the
figures, we have assumed an observation point located at
$\Phi=0^\circ$ and $\Theta = 45^\circ $ or $90^\circ$ (as indicated).


Direct visual inspection of the waveforms gives some indication of how
well the NK and TB waveforms match. In each of the cases illustrated,
the kludge waveforms manage to capture the overall wave pattern, and
for the orbits with $p>8M$ they almost exactly match the TB
waveforms. For ultra-relativistic orbits (e.g., with $p\lesssim 5M$
and $1-a/M\ll 1$), the finer structure in the TB waveforms is clearly
not reproduced by the quadrupole-octupole kludge waveforms. As we have
discussed, these features are the imprints of higher multipole
components in the radiation which are amplified by back-scattering,
and thus are not expected to appear in the NK waveforms.  Waveforms
generated using the Press formula (not shown) do have some finer
features due to the presence of higher multipole components, as we
might have hoped, but they are not nearly as complicated as the
structure of the TB waveforms.

One also notices that for certain parts of the waveforms (e.g.,
Fig.~\ref{circ3}), there is a disagreement in the amplitude while the
phase is accurately reproduced. This amplitude discrepancy is
periodic, i.e., the points where the amplitude is poorly reproduced
occur at regular intervals. This suggests that the kludge waveforms
are missing some periodic components, as we might expect since they
are truncated expansions in multipole moments. Indeed, the amplitude
discrepancy is less pronounced in the Press waveform, which includes
contributions at all multipoles. For the purposes of data analysis, it
is important to have templates with a high overlap with the true
signals, and to achieve that it is much more important that the
waveform phase is reproduced rather than the waveform amplitude. The
waveform phase is determined by the orbit generating the gravitational
radiation. The fact that the kludge waveforms are based on true
geodesic orbits is presumably the reason that we find, {\it post
facto}, such impressively high overlaps with TB waveforms and
especially for those from circular-inclined orbits.

A comprehensive set of data for the overlaps between NK and TB
waveforms is given in Tables~\ref{o-lap_equat}, \ref{o-lap_circ} and
\ref{o-lap_generic} in the Appendix. These were computed using the
overlap function described in Section~\ref{overlaps} and assuming a
central black hole mass of $M=10^6 M_{\odot}$. This was chosen since
preliminary event rate estimates suggest the inspirals of $\sim
10M_{\odot}$ BHs into $\sim 10^6 M_{\odot}$ SMBHs will dominate the
LISA detection rate \cite{rates}. These tables indicate that if the
orbital periastron $r_p \gtrsim 5 M$, the overlap between TB waveforms
and both the quadrupole-octupole and Press waveforms stays above
$\sim0.95$. We also find that both these expressions have better
performance than the pure quadrupole waveforms \erf{quad}, but there
is little difference between quadrupole-octupole and Press
waveforms. The kludge and TB waveforms begin to deviate significantly
for strong-field, ultra-relativistic orbits with $ r_p \lesssim 4M$,
with the overlap dropping to $\sim 50\% $ for orbits that come very
close to the horizon. Disappointingly, the Press waveforms do not seem
to do much better than the quadrupole-octupole waveforms in this
strong field regime, despite the inclusion of additional multipole
components. The Press waveforms include only the ``direct'' contribution to the higher multipoles, i.e., the piece that arises from fast motion in flat spacetime. In true EMRI GWs, the higher multipole contributions are significantly enhanced by ``tail'' effects, i.e., the backscattering of radiation from the background geometry. Since we are using the Press formula in flat space, we do not include this back-scattering enhancement. This is presumably why using the kludge Press waveforms does not significantly improve the overlap with TB waveforms. The Press waveforms do perform consistently better for circular orbits and weak-field eccentric orbits, but the difference
between the two approaches is usually small (with the exception of a
few cases which we discuss in more detail later). We conclude that the
quadrupole-octupole waveform model is sufficient and there is not much
gain from using the computationally more intensive Press formula.

To summarize, we find that kludge waveforms are accurate --- and very
quick to generate --- substitutes for TB waveforms for all orbits with
periapse $r_p \gtrsim 5M$. In Schwarzschild, the separatrix between
bound and plunging orbits is at $(r_p)_S = (6 + 2e)/(1+e) \,M$, so the
kludge waveforms will be accurate throughout any inspiral with
eccentricity at plunge $e \lesssim 1/3$. Computations of inspirals
into Schwarzschild black holes \cite{GHK2} indicate that, in many
cases, the residual eccentricity at plunge will be small, so that
kludge waveforms will be suitable for the majority of Schwarzschild
inspirals.  For retrograde orbits around Kerr black holes ($ 90^\circ
\leq \iota \leq 180^\circ$), the minimal periapse is at even larger
radii, so that even weaker restrictions must be imposed on the
eccentricity at plunge for the kludge to be valid. In contrast, for
prograde orbits ($ 0^\circ \leq \iota \leq 90^\circ$), an increased
black hole spin allows stable orbits to exist much deeper in the
strong field. For $a=0.9M$, the separatrix for circular orbits is at
$r_p \approx 2.3M, 2.6M, 3.7M$ for orbits with inclinations of
$0^\circ, 30^\circ, 60^\circ$ respectively; for orbits with a plunge
eccentricity of $e=1/3$, it is at $r_p\approx2.0M, 2.2M, 3.1M$ for the
same inclinations. As $a \to M$, the separatrix of equatorial orbits
decreases even more, asymptotically approaching $(r_p)^{\rm pro}_K
= M$ \cite{zoom}. Kludge waveforms are not particularly good in this
regime, with overlaps $\sim 50\%$; fortunately, this corresponds to a
comparatively small region of parameter space.  The evolution proceeds
through this region very quickly, so we do not lose much
signal-to-noise ratio by failing to match the waveforms in this
region.

If the overlap between a given signal and the best-fit template in a
search bank is less than $1$, this leads to a decrease in the maximum
distance to which that signal can be detected, and a corresponding
reduction in event rate. For the purposes of detection, overlaps as
low as $50\%$ might well be considered good enough, if the
astrophysical event rate is sufficiently large \cite{rates}. However,
for the purposes of parameter estimation, much higher overlaps ($\gsim
95 \%$) will be required in general. It is clear from the results in
this paper that this is only partially achievable by the existing
family of kludge waveforms. Nonetheless, these waveforms might be
useful for LISA data analysis as search or detection templates over
some (perhaps a large part) of the astrophysically relevant portion of
the $\{a/M, p, e, \iota \}$ parameter space. The waveforms may also
provide sufficiently accurate estimation of the source parameters (in
certain regions of parameter space) that they could be used as the
first stage in a hierarchical search. The purpose of such a search
would be to identify ``interesting'' regions of parameter space for
follow up with more accurate waveforms. This is the main conclusion of
this paper.

However, the regions where kludge waveforms are good enough must be
identified more carefully by comparison to accurate {\it inspiral}
waveforms. As we have discussed earlier, the flat space emission
formula used in the construction of the NK waveforms ignores all
effects of scattering from the background curvature. These ``tail''
terms make a significant contribution to the waveform structure, and
build up over the course of an inspiral. Although we have found good
overlaps with geodesic waveforms and circular-inclined inspiral
waveforms (see Section~\ref{inspcomp}), comparisons to inspirals of
eccentric-inclined orbits are required to properly assess the
importance of including the tail terms.  Accurate, self-force
waveforms for such orbits will not be available for a few years and
only then will it be possible to firmly demarcate the regime of
usefulness of the present, or further improved, NK waveforms.


\subsubsection{Frequency domain comparison}

\begin{figure}
\includegraphics[height=0.33\textheight, keepaspectratio=true,
angle=0]{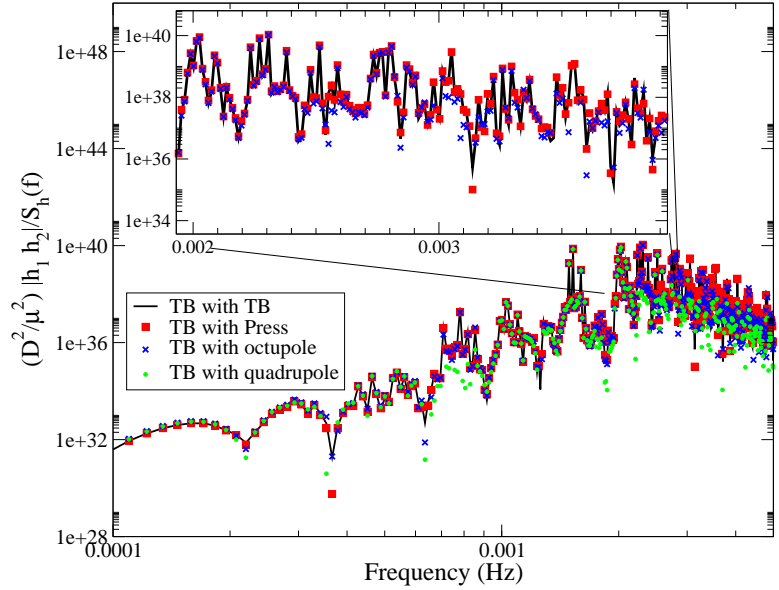}
\caption{Comparison of the integrand of the SNR in the frequency domain
$[\tilde{h}(f)\tilde{s}^*(f)]/S_h(f)$ where $\tilde{h}(f)$ is determined from the
quadrupole, quadrupole-octupole and Press waveforms, and $s(f)$
is determined from the TB waveform. We have chosen orbital parameters $p=
12M,~e=0.1,~\iota = 120^\circ $ and spin $a=0.9M $. One can see that
Press waveform performs better than quadrupole-octupole waveform at
high frequencies.
}
\label{NKs}
\end{figure}

To better understand the overlaps quoted in the tables we consider
here the integrand of Eq.\ (\ref{max_t}) for the value of $t$ which
maximizes the overlap:
\begin{equation}
\frac{\tilde{x}(f)\tilde{h}^*(f) + \tilde{x}^*(f)
\tilde{h}(f)}{S_h(f)}e^{i2\pi ft}.
\label{integrand}
\end{equation}
In Figure~\ref{NKs} we plot this function for the generic orbit
$p=12M,~e=0.1,~\iota=120^\circ$ and $a=0.9M$.  For the signal
$\tilde{x}(f)$ we use the TB waveform and correlate it with templates
$\tilde{h}(f)$. As templates we use the TB waveform itself (black
solid line), the Press waveform (squares), the quadrupole-octupole NK
waveform (crosses) and the quadrupole NK waveform (circles). We have
deliberately chosen a case for which including higher harmonics
significantly ($>10\% $) improves the overlap, especially in the
presence of white dwarf confusion noise (see Table
\ref{o-lap_generic}).

One can see that the main contributions to the overlap come from
several dominant harmonics. For a circular equatorial orbit the main
harmonic would correspond to twice the orbital frequency, but in
general the main harmonics depend on eccentricity \cite{AK}.  Besides
the harmonics coming from the azimuthal motion there are many
additional components coming from beating between harmonics of the
fundamental frequencies of the $\phi$-, $\theta$- and $r$-motion:
$\Omega_{\phi}, \Omega_{\theta}, \Omega_{r}$ (see \cite{DrascoHughes}
for a Fourier decomposition of the orbital motion).  Depending on the
mass $M$ of the central BH, the dominant harmonics may lie in the most
sensitive part of LISA's frequency range or they can be suppressed by
confusion noise. In the latter case the higher harmonics, although
smaller in amplitude, are effectively enhanced by the inverse power
spectral density and therefore play an important role in SNR
accumulation. Looking at the inlaid box in Figure~\ref{NKs} we can
compare the quadrupole-octupole and Press waveforms at high
frequency. Here, the Press waveform does perform better \--- e.g.,
around 3mHz the quadrupole-octupole NK waveform has failed to
reproduce some harmonics in the TB signal, whereas the Press waveform
does so pretty well. We will discuss the frequency representation of
the signals further in the next subsection.

\subsection{Comparison to Teukolsky-based inspiral waveforms}
\label{inspcomp}

As described earlier, most of the existing TB waveforms describe GW
emission from geodesic orbits, which is why we have focused on these
for the comparisons above. However, TB waveforms have also been
constructed for the inspiral of circular-inclined orbits
\cite{scott_insp}. To illustrate the applicability of the kludge to
the detection of realistic, inspiralling EMRI waveforms, we here
compare a kludge inspiral waveform to a TB circular-inclined inspiral
waveform generated using the method described in
\cite{scott_insp}. The kludge inspiral trajectory is constructed using
the flux formulae described in \cite{GHK2}, and then the particle
trajectory is computed as discussed in Section~\ref{insptraj}. We use
the quadrupole-octupole formula \erf{octup} to generate a kludge
waveform from this trajectory.

\begin{figure}[ht]
\includegraphics[height=0.33\textheight, keepaspectratio=true,
angle=0]{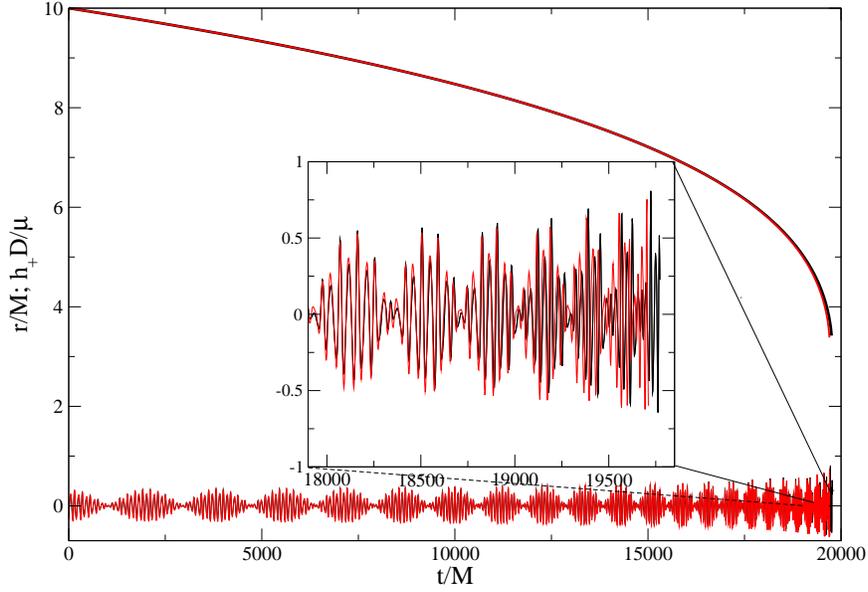}
\caption{The orbital evolution, $r(t)$, and $h_{+}$ polarization of
the GW signal, for an inspiralling quasi-circular inclined orbit with
initial parameters $p = 10 M,~ e=0,~\iota = 45^\circ$ and
$a=0.9M,~\mu/M = 10^{-2},~\Theta=90^\circ$. The black curve
corresponds to the TB waveform and the red (grey in b\&w version) line represents the
quadrupole-octupole NK waveform. The inlaid box shows the part of the
waveform where the TB and NK results start to deviate, which
correlates with the deviation in the inspiral.}
\label{InspCom}
\end{figure}

A visual comparison is shown in Figure~\ref{InspCom}. There we plot
both the inspiral trajectory $r(t)$ and the gravitational waveform
$h_{+} $. One can see an excellent agreement up to the last several
cycles, where we would not trust either type of waveform anyway.  We
have used a large mass ratio ($10^{-2}$) to compute the waveforms in
order to speed up the orbital evolution. We should point out that the
comparison is not improved if we use Press waveforms instead of
quadrupole-octupole waveforms.

\begin{figure}[ht]
\includegraphics[height=0.33\textheight, keepaspectratio=true,
angle=0]{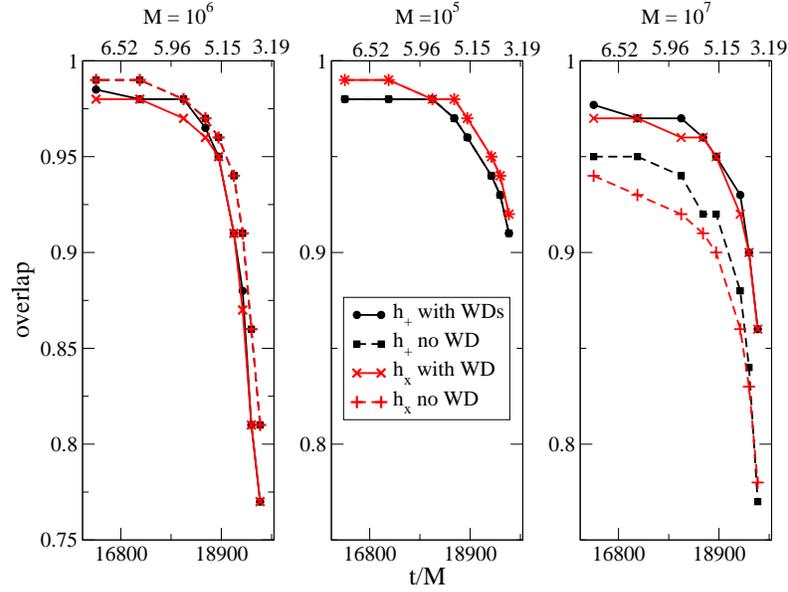}
\caption{Overlap between TB and NK waveforms ($h_{+}$ and 
$h_{\times}$) for an inspiralling
(quasi)circular-inclined orbit as a function of truncation time.  We
show overlaps for different masses of the central BH: $M= 10^5, 10^6,
10^7 M_{\odot}$ and for sensitivity curves with and without white dwarf (WD) confusion noise. One can see that the overlap drops as 
we increase the truncation time and the mass $M$. On the top horizontal axis we show the orbital radius $r/M$ corresponding to
the truncation time.
For $M=10^5M_{\odot}$ the lines with and without WD confusion lie on 
top of each other.}
\label{InspOlap}
\end{figure}

We may quantify the agreement in the waveforms by again using the
overlap function, which we plot as a function of truncation time in
Figure~\ref{InspOlap} for three different masses of the central black
hole. One can see that for the majority of the inspiral we obtain
quite high overlap, but it decreases when we include the final stages
of the inspiral. For the inspiral in Figure~\ref{InspCom}, the overlap
(when $M=10^6M_{\odot}$ and in presence of the white dwarf background)
up until radii of $r = 8M, 6M, 5M, 4M, 3.5M$ is $0.985, 0.974, 0.92,
0.79, 0.755$ respectively. The overlap over the whole inspiral is
still greater than $75\%$. An overlap of $75\%$ corresponds to a
factor of $\sim 2$ reduction in event rate. However, overlaps of
$90\%$ or $95\%$ correspond to a reduction in event rate of only $1.4$
or $1.17$ respectively. A significant amount of signal to noise ratio
accumulates over the early stage of inspiral, and so it may well be
that kludge inspiral waveforms can be used to efficiently detect and
track events until they reach $r\approx5-6M$. Alternative techniques
could then be employed for detection during the latter stages of
inspiral.

\begin{figure}[ht]
\includegraphics[height=0.33\textheight, keepaspectratio=true,
angle=0]{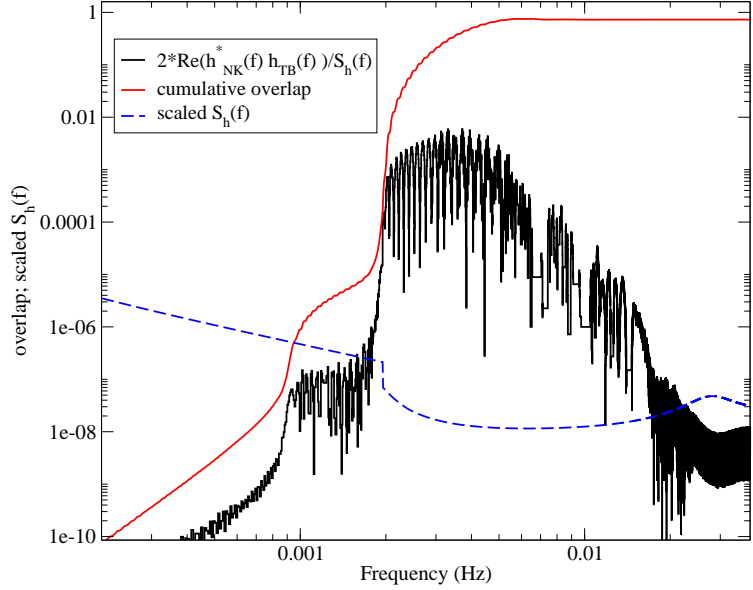}
\caption{The integrand of the overlap between TB and NK waveforms in
the frequency domain (solid black line).  Breaks in the curve
correspond to a negative correlation. We also overplot the
accumulative overlap (red solid line) and scaled $S_{h}(f)$ as a
dashed blue line.}
\label{InspInt}
\end{figure}

The mass of the central BH determines the frequency range through
which the inspiral occurs. One could also evaluate the spectrum of the
signal as a function of the dimensionless frequency $(Mf)$ \--- the
effect of changing $M$ is then to shift the PSD $S_h(f)$ right as $M$
is increased or shift it left as $M$ is decreased.  In
Figure~\ref{InspInt} we again plot the overlap integrand
(\ref{integrand}) for $M=10^6M_{\odot}$. We also overplot the
cumulative overlap
$$
2 \int_0^{F}\frac{\tilde{x}(f)\tilde{h}^*(f) + \tilde{x}^*(f)
\tilde{h}(f)}{S_h(f)} df
$$
and a scaled PSD. By shifting $S_h(f)$ on the plot left/right one can
see how the central mass affects the amount each harmonic contributes to the SNR/overlap. The
higher harmonics are modeled less accurately by NK waveforms and this
leads to an overall drop in the overlap for
$M=10^7M_{\odot}$. However, the majority of the high frequency
contribution to the waveforms comes from the end of the inspiral which
we do not reproduce very accurately, rather than from higher harmonics
emitted during the earlier, moderately relativistic part of the orbit.

We should also mention that there is a problem with spectral leakage
\cite{Percival} (due to the large dynamic range) in the estimation of
GW spectra. This is not a big problem on its own but for high $M$ this
leakage (at high frequencies) is amplified by the inverse power
spectral density and leads to an erroneous result.  In order to reduce
this effect we have applied a window function
$$
w(t) = \frac1{2}\left( 1 + \tanh\left[ A (t-\kappa T)\right] + \tanh\left[ A(T - t - \kappa T) \right]
\right)
$$
to the template and signal for $M=10^7M_{\odot}$ and in the presence
of confusion noise.  The parameters $\kappa$ and $A$ govern the
steepness and cut-off point of the window; $T$ is the waveform's
duration. This function reduces the effect of the end of inspiral on
the overlap and leads to an artificial increase in the overlap by a
few percent.

\begin{figure}[ht]
\includegraphics[height=0.27\textheight, keepaspectratio=true,
angle=0]{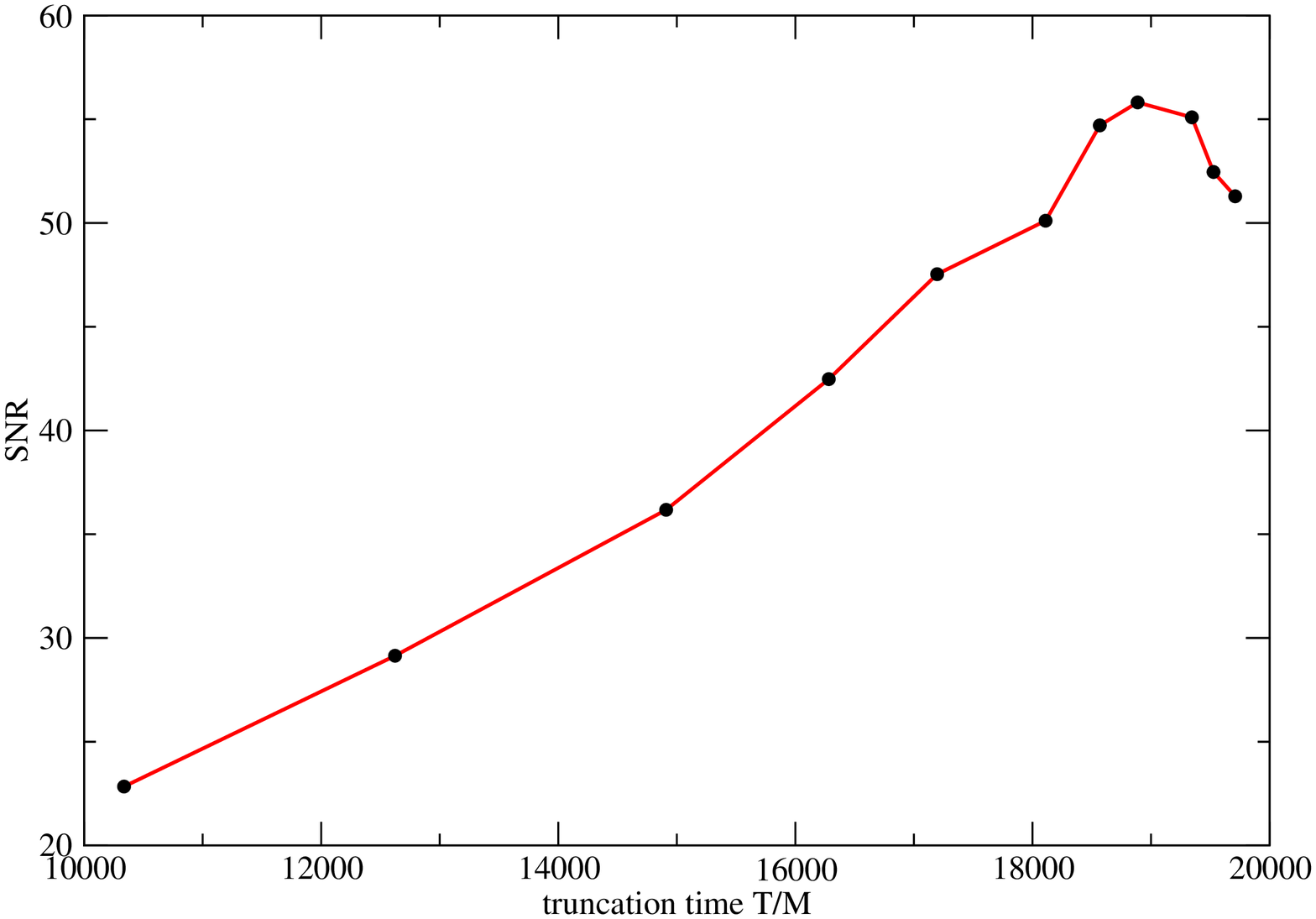}
\caption{In this plot we present SNR obtained by applying the truncated NK waveform as a template to the full inspiral represented 
by TB waveform: 
$SNR(h^{NK}_{tr}, s^{TB}) = \frac{(h^{NK}_{tr}|s^{TB})}{(h^{NK}_{tr}|h^{NK}_{tr})^{1/2}}$ normalised by the total SNR:
$SNR(s^{TB}, s^{TB}) = (s^{TB}|s^{TB})^{1/2}$. The primary mass was taken to be $10^6 M_{\odot}$ and the sensitivity curved contained WD confusion noise. The total SNR for the source at 10 Gpc is approximately 66 (high SNR is due to large ($10^{-2}$) mass ratio).}
\label{InspSNR}
\end{figure}

As mentioned above, a significant amount of signal to noise ratio
accumulates over the early stages of the inspiral. To illustrate this we show in Figure~\ref{InspSNR} the accumulation of SNR as a function of time. To generate this plot, we have assumed the TB waveform is the signal 
and have used the NK waveform, truncated at different times, as the search template.
One can see that we can recover up to 85\% of the maximum SNR (i.e., the SNR if the template and signal were identical). Including the last few waveform cycles we see a drop in the SNR ($\sim 8$\%) due to the mismatch in the waveforms at the end of the inspiral.

Employing Press waveforms helps to improve overlaps further, but for
ultra-relativistic orbits the omitted tail contribution to the
waveform becomes increasingly significant.

\subsection{Comparison to Teukolsky-based and PN fluxes}
\label{results_flux}

It is possible to construct expressions for the flux of energy and
angular momentum carried by the kludge GWs, and use these as estimates
of the energy and angular momentum lost from the orbit in a true
inspiral. The energy and angular momentum content of a TT GW field,
$h^{jk}_{TT}$, propagating in flat space at large distances from the
source is determined from the Isaacson energy momentum tensor of the
wave field \cite{isaacsona,isaacsonb}
\begin{equation}
T_{GW}^{\mu\nu} = \frac{1}{32\pi} \left< h_{TT}^{jk,\mu}
h_{TT}^{jk,\nu} \right>.
\label{GWemtensor}
\end{equation}
Integration of this expression gives the energy and angular momentum
loss rates due to GW emission (equations ($4.13$) and ($4.22'$) of
\cite{thorne80}) as
\begin{eqnarray}
\dot{E} &=& \frac{1}{16\pi} \int \left< h_{TT}^{jk,t} h_{TT}^{jk,t}
\right> r^2 \, \rmd \Omega \label{genEdot}
\\
\dot{L}_{i} &=& \frac{1}{16\pi} \int
\left<\epsilon_{ipq}h_{TT}^{pa}h_{TT}^{aq,t} -\frac{1}{2}
\epsilon_{ipq} x_{p}h_{TT}^{ab,q} h_{TT}^{ab,t} \right> r^{2} \rmd
\Omega \label{genJdot}
\end{eqnarray}
Using the quadrupole-octupole formula to generate $h^{jk}_{TT}$, these
expressions may be written in terms of the multipole moments of the
source as
\begin{eqnarray}
\langle \dot{E} \rangle &=& -\frac{1}{5} \left< {I^{jk}_{TT}}^{(3)}
{I^{jk}_{TT}}^{(3)} +\frac{5}{189} {M^{jkl}_{TT}}^{(4)}
{M^{jkl}_{TT}}^{(4)} +\frac{16}{9} {J^{jk}_{TT}}^{(3)}
{J^{jk}_{TT}}^{(3)} \right> \label{improvEdot}
\\
\langle \dot{L}_{i} \rangle &=& -\frac{2}{5} \; \epsilon^{ikl} \left<
{I^{ka}_{TT}}^{(2)} \; {I^{al}_{TT}}^{(3)} +\frac{5}{126}
{M^{kab}_{TT}}^{(3)} {M^{lab}_{TT}}^{(4)} +\frac{16}{9}
{J^{ka}_{TT}}^{(2)} {J^{qa}_{TT}}^{(3)} \right> \label{improvJdot}
\end{eqnarray}
In these expressions $J_{TT}^{ij} = \epsilon_{jlm}S_{TT}^{mli}$ and
bracketed numbers in superscripts denote the number of time
derivatives, e.g., ${I^{ka}_{TT}}^{(2)} = \rmd^2 (I^{ka}_{TT})/\rmd
t^2$. If the kludge waveform is computed using the quadrupole formula
\erf{quad}, only the leading order $I^{jk}_{TT}$ terms remain. If a
Press waveform is used, there is an infinite sum of multipole
components, and therefore expressions (\ref{genEdot}--\ref{genJdot})
must be employed directly.

\begin{figure}[ht]
\includegraphics[height=0.33\textheight,
keepaspectratio=true,angle=270]{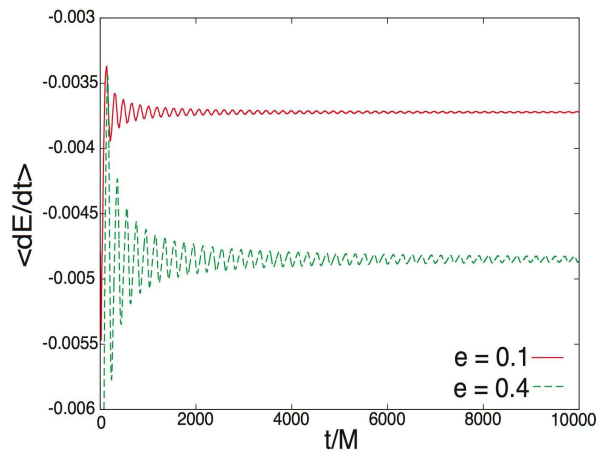}
\caption{Convergence of the averaged energy flux $\langle \dot{E}
\rangle $ with respect to the averaging time $T$ for a pair of
equatorial-eccentric orbits. Note the increased convergence rate with
decreasing eccentricity.}
\label{fig1}
\end{figure}

The angle brackets in expressions (\ref{genEdot}--\ref{improvJdot})
mean ``average over several gravitational wavelengths''. This is
achieved by taking a time average of the instantaneous flux expression
inside the angle brackets, i.e.,
\begin{equation}
<X> = \frac{1}{T} \int_{0}^T X \,dt
\label{tav}
\end{equation}
where $T$ is an appropriate averaging time. It is important to ensure
that this averaging time is long enough. For circular equatorial
orbits, the symmetries of the system ensure that the instantaneous and
averaged fluxes are identical. For an eccentric equatorial orbit,
there is only one periodicity in the flux --- the period of the radial
motion. Thus integrating over one radial period is sufficient to give
an accurate average flux. For generic orbits, the averaging is more
complicated since the periods of the motion in the radial and $\theta$
directions are in general different and non-commensurate. The
averaging time therefore has to be long enough to encompass several
periods of both motions. We illustrate this in Fig.~\ref{fig1} by
plotting $\dot{E}$ as function of $T$ for a generic orbit with
parameters $p = 4M, a=0.99M, \iota=60^{\circ}$ and $e=0.1, 0.4$.  The
flux has been computed using the pure quadrupole waveforms
\erf{quad}. It is clear from Fig.~\ref{fig1} that the averaged flux
converges over time, and this convergence is more rapid for the lower
eccentricity orbit. In all subsequent flux calculations, we took
$T=10^5M$ (which corresponds to $\sim 6$ days for $M = 10^6 M_\odot$).

\begin{figure}
\includegraphics[height=0.33\textheight, keepaspectratio=true,
angle=270]{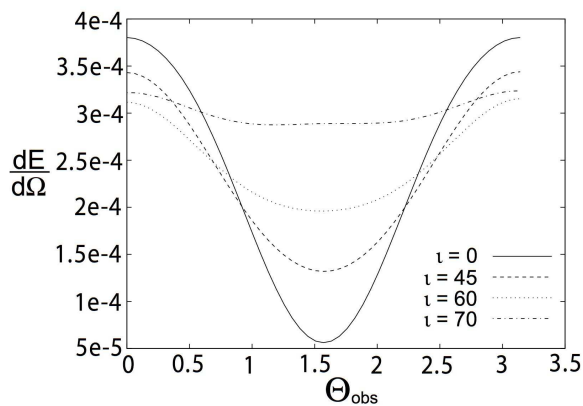}
\caption{Angular radiation pattern \--- energy radiated per unit solid
angle as a function of the colatitude of the observer, $\Theta_{obs}$,
for orbits with $p=5M$, $e=0.4$, $a=0.9M$ and a sequence of
inclinations, as labelled.}
\label{fig2}
\end{figure}

In Figure~\ref{fig2} we show the angular distribution of the
gravitational radiation from an EMRI orbit of given $p$ and $e$, but
for several different orbital inclinations, $\iota$. This picture is
more or less typical of the majority of orbits. The variation of the
energy flux with the sky position of the observer is indicative of
beaming. This is strongest for equatorial orbits and reduces as the
orbital inclination is increased. Equatorial orbits are restricted to
a single plane, while inclined orbits wander through more of the
spacetime. This wandering averages out the beaming over the sky. The
larger the orbital inclination, the more of the phase space the body
explores, the more averaging occurs and the more homogeneous the sky
distribution of the energy flux.

In Tables~\ref{tab3} \& \ref{tab4} we show flux data for a series of
orbits and black hole spins.  We tabulate the flux computed via
solution of the Teukolsky equation, and fluxes computed using
quadrupole kludge waveforms and equation \erf{improvEdot}. We quote
only the quadrupole results since the difference from using either the
quadrupole-octupole or Press waveforms is quite small.  As we
mentioned in Section~\ref{inventory}, Tanaka et.al.\ \cite{shibata}
and more recently Gair et al.\ \cite{GKL} computed approximate fluxes
for Schwarzschild orbits based on the quadrupole waveform
approximation \erf{quad}. For a random choice of orbits we have found
excellent agreement with their results.

For further comparison, we also tabulate the ``kludge'' fluxes that
have been used to construct inspirals and inspiral waveforms for
scoping out LISA data analysis. The difference between these and our
current results is a measure of the inconsistency in the kludged
inspiral waveforms. In constructing an inspiral waveform, one set of
expressions is used to evolve the orbit (given in \cite{GHK,GHK2}),
and a different prescription (the one described here) is used to
generate the waveform. This means that the energy carried by the
kludge GWs is not equal to the energy lost by the orbit that is
supposedly emitting these GWs. This leads to inaccuracies when the
kludge waveforms are used to compute SNRs \cite{rates}. In the
original prescription for kludge inspirals \cite{GHK}, the phase space
trajectories were generated using the leading order post-Newtonian
results of Ryan \cite{ryan}, evaluated for relativistic orbital
parameters. Ryan's expression for the energy flux is
\begin{equation}
\dot{E}= -\frac{32}{5} \frac{\mu^2}{M^2} \left(\frac{M}{p}\right)^5
(1-e^2)^{3/2}\left [ f_1(e) -\frac{a}{M}\left(\frac{M}{p}\right)^{3/2} f_2(e)
\cos\iota \right]
\label{GHKEdot}
\end{equation}
where the $e$-dependent coefficients are,
\begin{equation}
f_1(e) = 1 + \frac{73}{24} e^2  + \frac{37}{96}e^4, \quad\quad\quad
f_2(e)= \frac{73}{12} + \frac{823}{24} e^2 + \frac{949}{32}e^4
+ \frac{491}{192}e^6 .
\end{equation}
In a more recent version of the kludge \cite{GHK2}, this prescription
for the evolution of the orbital parameters was improved. The new
expression for $\dot{E}$ is complicated and we do not quote it here,
but we also tabulate those results in Tables~\ref{tab3} \& \ref{tab4},
under the heading ``GG''.

We note that in true black hole spacetimes, energy and angular
momentum are lost into the horizon as well as to infinity. The
Teukolsky equation provides a prediction for both the flux of
radiation at infinity and the flux down the black hole horizon.  When
using Eqs.\ \erf{genEdot}--\erf{improvJdot}, we are only computing the
energy and angular momentum carried to infinity in the kludge
GWs. This should therefore only be compared to the Teukolsky flux at
infinity. The discrepancy between these numbers is then a measure of
the error associated with using the kludge waveforms to estimate
signal to noise ratios for true inspirals.  The TB fluxes quoted in
the tables are accordingly just the infinity piece of the fluxes. When
computing inspirals, which was the purpose for which the ``GG''
expressions were derived, the energy and angular momentum lost by the
orbit should include the horizon piece of the flux. This should be
born in mind when considering the flux tables. However, the relative
contribution of the horizon flux is only significant in the extremely
strong field regime where both the ``GG'' expressions and the kludge
fluxes cease to be valid.

In general, we find that the fluxes computed from the kludge waveforms
agree less well with Teukolsky results than do the ``GG'' fluxes. The
kludge fluxes do improve over ``GG'' results for some retrograde
orbits, and for some orbits with high eccentricity, but these are
orbits for which the ``GG'' expressions are known to lack some
terms. Once the ``GG'' expressions are further improved along the
lines discussed in \cite{GHK2}, it is likely that the ``GG'' fluxes
will always be closer to TB results. Typically, the fluxes computed
from the kludge waveforms underestimate the actual energy and angular
momentum lost by the orbit, and also underestimate the ``GG''
fluxes. This suggests that inspiral waveforms that combine the two
prescriptions will carry too little energy to infinity for the
evolution seen in the orbit (or equivalently, the inspiral will
proceed too quickly for the energy being carried by the GWs). In \cite{GKL}, a similar
comparison was performed for inspirals into Schwarzschild black holes
and they found that the kludge GWs contained {\it too much} energy, in
contrast to this result. This difference arises primarily because, in
that paper, inspirals were evolved using the ``Ryan'' fluxes, rather
than the more up to date ``GG'' fluxes. The signal to noise ratio squared of a monochromatic source contributed when the frequency is in the range $f \rightarrow f+{\rm d}f$ is proportional to $\dot{E} {\rm d}f/(f^2 \, \dot{f}\,S_h(f))$. The kludge waveforms get $\dot{f}$ largely correct (otherwise the high overlaps would not be maintained), but the error in $\dot{E}$, arising from an amplitude discrepancy, will lead to an error in SNRs computed using kludge inspiral waveforms. Since $\dot{E}$ is in general an underestimate, the kludge SNRs will probably be underestimates of the true SNRs. The discrepancy in $\dot{E}$ is at most a few tens of percent. A discrepancy of $25\%$ {\it throughout} the inspiral would lead to an error in the SNR of only $13\%$. However, the actual error will be much less, since for the majority of the inspiral the kludge $\dot{E}$ is much closer to the true value, and, as we saw earlier, most of the SNR comes from the early inspiral. The kludge waveforms are thus accurate enough to be used for SNR calculations to estimate astrophysical event rates \cite{rates}, in which the astrophysical uncertainties far outweigh the waveform uncertainties. However, we must emphasize that amplitude discrepancies of this sort do not limit the applicability of the kludge waveforms for data analysis, provided the overlaps with true waveforms are high. If the kludge waveforms were used for parameter extraction, such discrepancies would lead to an error in the distance estimate only. In practice, we are likely to use kludge waveforms for {\it detection} rather than source characterization, and for that the waveform amplitude is irrelevant.

\begin{figure}[ht]
\includegraphics[height=0.3\textheight,keepaspectratio=true,angle=0]{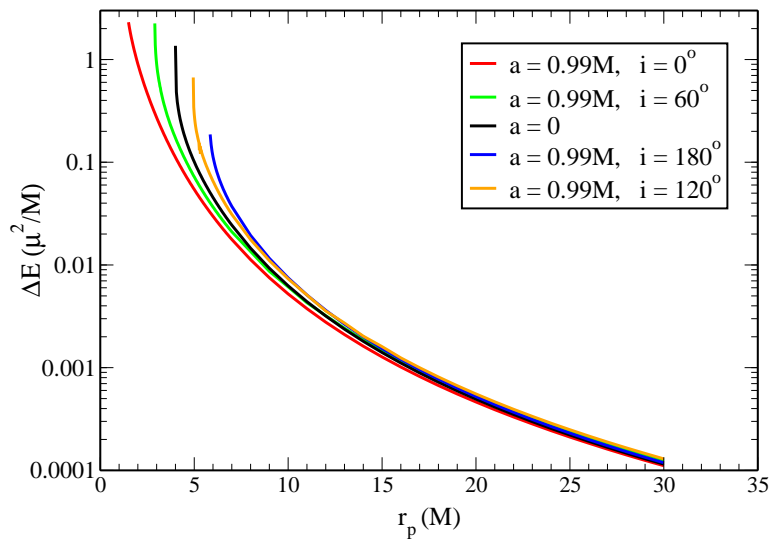}
\caption{Total radiated energy for Kerr parabolic orbits; as function
of periastron.}
\label{parfig}
\end{figure}

Table~\ref{tab4} tabulates data for ``parabolic'' orbits, $e=1$. The
GW emission from such orbits is important for the capture problem,
i.e., the mechanism by which compact objects are put onto EMRI
orbits. Gravitational radiation from such orbits has been studied in
the past using the Teukolsky formalism \cite{detweiler},\cite{kojima},
but tabulated data has only appeared in a recent time-domain analysis
by Martel \cite{martel}. Those results are claimed to be accurate at
the level of $ \lesssim 1 \% $, and therefore can be treated as exact
for our purposes. Table~\ref{tab4} contains the relevant data together
with our kludge results. For a periastron $ r_p \sim 5 M$ the error in
using the kludge fluxes is $\sim 30 \% $ and rapidly decays (grows) as
we move to larger (smaller) periapse. In this Table we have also
included some data for non-equatorial parabolic orbits, but for this
class of orbits there are as yet no available TB results. Our
experience with bound generic orbits suggests it is likely these data
will have accuracies of the same order as the equatorial orbits with a
similar periastron. In Fig.~\ref{parfig}, we plot $\Delta E$ as a
function of $r_p$ and $L_z$ (left and right panels, respectively) for
parabolic orbits with $a/M = 0,~0.99$ and $\iota = 0^\circ ,~60^\circ
,~120^\circ ,~180^\circ$. All curves are computed using the quadrupole
waveform flux \erf{quad} and are terminated at the point where the
orbit plunges.
Fig.~\ref{parfig} has astrophysical significance as it displays
(within the accuracy of the present calculation) the amount of energy
lost in a single parabolic (or $e \approx 1$) encounter with a SMBH,
for a variety of inclinations and for a range of periastra that are
appropriate to the capture problem \cite{freitag}, \cite{GKL2}. This energy loss
data can be used to estimate the capture rate of compact objects by a
single massive black hole.

We note that it is clear from Table~\ref{tab4} that the ``GG'' results perform quite poorly for parabolic orbits. This is because the ``GG'' results are built on a small eccentricity expansion, which is no longer valid when $e=1$. At present the kludge fluxes are closer to TB results than the ``GG'' fluxes where comparisons can be made, and most likely are better approximations for the other orbits as well. The performace of the ``GG'' fluxes for highly eccentric orbits is now being improved, and it is likely that the next generation prescription will be able to approximate TB fluxes accurately for orbits of all eccentricities.

\section{Inclusion of conservative corrections}
\label{cons_pieces}

In the preceding sections, we have shown that kludge waveforms do very
well at reproducing the GW emission from geodesic orbits, and, if
coupled to an accurate prescription for the radiation fluxes, can very
accurately reproduce the GW emission from inspirals. However, the
radiation fluxes are not the whole story, as they represent only the
dissipative piece of the self-force. In true inspirals, the self-force
also has a ``conservative'' component.  This does not lead to
evolution of the orbital parameters and hence cannot be determined
from the flux of energy, angular momentum and Carter constant, but
does lead to a change in the phasing of the gravitational
waveform. There has been some debate recently about the importance of
including the conservative pieces of the self-force when constructing
templates for EMRI source detection \cite{drasco2,pound}. It is quite
possible that these corrections will be needed for detection, at least
for some range of source parameters; these terms will certainly be
needed for parameter determination. We aim to demonstrate here how
easily conservative terms can be included in the kludge waveform
prescription.

The conservative part of the self-force will include an oscillatory
component, which averages to zero, and a secular piece that leads to
accumulation of a phase error over time. The net effect of the secular
piece is that, for a given set of orbital parameters ($p$, $e$ and
$\iota$), the average frequencies of the motion in the $r$, $\theta$
and $\phi$ directions will differ from the expected values for a Kerr
geodesic.  The conservative self-force might also affect the harmonic
structure of the GWs, but the dominant contribution will come from
orbital dephasing.  In the spirit of the kludge, we will now try to
account for this part of the effect.

The effect of the conservative self-force can be characterized by the
changes in the $r$, $\theta$ and $\phi$ frequencies, averaged over
many orbits. We denote these frequency changes by $\delta \Omega_r$,
$\delta \Omega_{\theta}$ and $\delta \Omega_{\phi}$. Conservative
effects can then be included in the kludge orbital determination
simply by amending the evolution equations for the angular variables
$\psi$, $\chi$ and $\phi$ given earlier (Eqs.~\erf{dpsidt},
\erf{dchidt}, and~\erf{geoEqphi} respectively):
\begin{eqnarray}
\left(\frac{\rmd \psi}{\rmd t}\right) = \left(\frac{\rmd \psi}{\rmd
t}\right)_{\rm geo} + 2\pi\,\delta \Omega_r\;, \qquad
\left(\frac{\rmd \chi}{\rmd t}\right) = \left(\frac{\rmd \chi}{\rmd
t}\right)_{\rm geo} + 2\pi\,\delta \Omega_{\theta}\;, \qquad
\left(\frac{\rmd \phi}{\rmd t}\right) = \left(\frac{\rmd \phi}{\rmd
t}\right)_{\rm geo} + 2\pi\,\delta \Omega_{\phi}\;. \label{conscorr}
\end{eqnarray}
The subscript ``geo'' denotes expressions for the phase derivatives in
geodesic orbits given earlier. In general, the frequency shifts will
depend on the orbital parameters, and for an inspiralling source we
assume that these are evaluated for the instantaneous orbital
parameters, as discussed in Section~\ref{insptraj}.

To evaluate these frequency shifts in the framework of black hole
perturbation theory requires self-force calculations \cite{poissonLR}
which are not yet fully developed, but should be on the timescale of a
few years. Once the self-force corrections are known, it should be
possible to use them to compute the effective frequency shifts in the
framework we use here, i.e., objects moving on geodesics of the Kerr
spacetime, the parameters of which evolve with time. In the meantime,
conservative corrections are already known in the post-Newtonian
framework up to 3.5PN order \cite{3.5PN}.  These can be used to
compute leading order conservative corrections to include in the
kludge. We outline here a general method for such a calculation,
illustrated by the simplest case of a 1PN conservative correction to
circular orbits in the Schwarzschild spacetime. The generalisation to
3.5PN corrections for arbitrary orbits in Kerr will be presented in a
future paper.

The greatest difficulty in such a computation is to identify
coordinates between the post-Newtonian and perturbative formalism. The
former calculations tend to be carried out in harmonic coordinates
while the natural coordinate system to use for the latter are
black-hole centered coordinates, like the Boyer-Lindquist coordinate
system. The best approach is to use asymptotic observables,
specifically the orbital frequency, perihelion precession frequency,
orbital plane precession frequency and their first time derivatives,
to identify coordinates between the two formalisms and find the
missing conservative pieces. The power of this approach is that it
corrects the kludge in a physically meaningful way; however, this
comes at the cost of an effective $O(\eta)$ redefinition of the
orbital parameters, where $\eta = \mu/M $.  For extreme mass ratios,
$\eta < 10^{-4}$, this is less than the expected error in parameter
determination \cite{AK}.

The post-Newtonian model has two particles moving on orbits (expanded
to the stated post-Newtonian order) and evolving via post-Newtonian
radiation reaction expressions. These orbits include conservative
corrections to geodesic motion. For the case of circular orbits in
Schwarzschild we find at 1PN and with mass ratio corrections at linear
order only that the orbital frequency and its time
derivative are given by \cite{junker92}
\begin{eqnarray}
\frac{\rmd \phi}{\rmd t} &\equiv& \Omega =
\left(\frac{M}{R}\right)^{\frac{3}{2}} \,\frac{1}{M}\,
\left(1+\frac{\eta}{2} - \left(\frac{M}{R}\right) \left(\frac{3}{2} +
\frac{7}{4} \, \eta\right) \right) \label{PNOm} \\
\frac{\rmd \Omega}{\rmd t} &=& \frac{96}{5} \,
\frac{\eta}{M^2}\,\left(\frac{M}{R}\right)^{\frac{11}{2}}
\left(1+\frac{3}{2}\,\eta -
\left(\frac{M}{R}\right)\,\left(\frac{2591}{336}+\frac{13571}{672}\,\eta\right)\right)
\label{PNOmdot}
\end{eqnarray}
In the above, we use $M$ to denote the mass of the primary (not the
total mass as is conventional in the post-Newtonian approach), and $R$
to denote the post-Newtonian orbital semi-major axis. In the kludge,
the orbit is a geodesic of the Kerr spacetime, and the orbit is
evolved according to the prescription in \cite{GHK2}. For circular
equatorial orbits in Schwarzschild we have (at 1PN order)
\begin{eqnarray}
\Omega &=& \frac{1}{r}\,\left(\frac{M}{r}\right)^{\frac{1}{2}}
\label{KlOm}\\
\frac{\rmd r}{\rmd t} &=&
-\frac{64}{5}\,\frac{\mu}{M}\,\left(\frac{M}{r}\right)^{3}
\left(1-\frac{743}{336}\,\left(\frac{M}{r}\right)\right)
\label{Klrdot}\\
& & \Rightarrow \frac{\rmd \Omega}{\rmd t} =
\frac{96}{5}\,\frac{\eta}{M^2}\,\left(\frac{M}{r}\right)^{\frac{11}{2}}
\left(1-\frac{743}{336}\,\left(\frac{M}{r}\right) +
O\left(v^{\frac{3}{2}}\right)\right) \label{KlOmdot}
\end{eqnarray}
Here, $r$ denotes the Schwarzschild radial coordinate of the orbit. If
we write
\begin{equation}
r=R\,\left(1+\left(\frac{M}{R}\right)\,b_0 + \cdots + \eta\, c_0 +
\eta\,\left(\frac{M}{R}\right) c_1 +\cdots \right)
\end{equation}
and substitute into expression \erf{KlOm} or \erf{KlOmdot}, keeping
corrections at 1PN order and linear in the mass ratio, we can compare
to equations \erf{PNOm} or \erf{PNOmdot} and solve for $b_0$, $c_0$
and $c_1$. The comparison between \erf{KlOm} and \erf{PNOm} gives a
different result for $c_0$ and $c_1$ than the comparison between
\erf{KlOmdot} and \erf{PNOmdot}, since the kludge prescription ignores
conservative effects. We therefore write
\begin{equation}
\Omega = \frac{1}{r}\,\left(\frac{M}{r}\right)^{\frac{1}{2}} + \delta
\Omega = \frac{1}{r}\,\left(\frac{M}{r}\right)^{\frac{1}{2}} +
\eta\,\left(\frac{M}{r}\right)^{\frac{3}{2}}\,\frac{1}{M}\, \left(d_0
+ d_1\,\left(\frac{M}{r}\right)\right)
\label{Klpluscons}
\end{equation}
and solve simultaneously for $b_0$, $c_0$, $c_1$, $d_0$ and $d_1$
[bearing in mind that \erf{Klrdot} not \erf{KlOmdot} is the
fundamental quantity to examine for computing inspiral in the
kludge]. We find that the required parameters are
\begin{equation}
b_0 = 1, \qquad c_0 = -\frac{1}{4}, \qquad c_1 = \frac{845}{448},
\qquad d_0 = \frac{1}{8}, \qquad d_1 = \frac{1975}{896}\;.
\label{concorrparams}
\end{equation}
We note that the correction to $\Omega$ scales as $R^{-3/2}$, compared
to the scaling of $\dot{\Omega}$, which is $R^{-11/2}$. In the
weak-field, $R\rightarrow \infty$, the conservative correction will be
much more significant than the radiative correction. This is in
keeping with what Pound et al. \cite{pound} found for a particle
moving under an electromagnetic self-force.

Extension to generic orbits is straightforward.  In that case, we have
two extra observables, the perihelion precession frequency and the
rate of precession of the orbital plane, and two extra coordinate
dependent orbital parameters, the eccentricity and orbital
inclination. Identifying the rate of change of the orbital frequency,
the perihelion precession frequency and the orbital plane precession
frequency between the post-Newtonian and kludge approaches gives us a
relation between the two sets of coordinates. Comparing the values of
the frequencies then tells us which $\delta \Omega_{\phi}$,
$\delta\Omega_{\theta}$ and $\delta \Omega_{r}$ must be added into the
kludge. In all of this we have chosen implicitly to identify the two
masses, $M$ and $\mu$, and the spin, $a$, between the post-Newtonian
and kludge models.  To our minds, it makes more sense to hold
these parameters fixed between the two models, at the cost of
modifying the definitions of the orbital elements $p, e, \iota$.

Once self-force data is available, it will be possible to obtain a consistent solution for the conservative correction in Boyer-Lindquist like coordinates by using fits to the self-force results. In using equation \erf{Klrdot}, we have implicitly assumed the relationship between energy/angular momentum and the radial coordinate is unchanged by the self-force. However, in self-force calculations, the energy and angular momentum of circular orbits at a given radius do change as a result of conservative effects. This is merely a manifestation of the $O(\eta)$ redefinition of $r$ implicit in this approach that we mentioned earlier. When comparing to self-force calculations, it will be preferable to include this redefinition of energy as well as the conservative corrections to $\Omega$ so that $r$ maintains it's meaning. Ultimately, both approaches are equivalent. Another way to obtain $d_0$ and $d_1$ for equation \erf{Klpluscons} is by expanding $\rmd \Omega/\rmd t$ as a function of $\Omega$ in both the PN and kludge approaches. Matching various orders in this expansion gives $d_0$ and $d_1$ directly without having to simultaneously solve for the coordinate transformation. The function $\rmd\Omega/\rmd t(\Omega)$ is a GW observable, so this approach ensures that the kludge will have the correct leading order form for this observable. However, this is a series truncated at a given PN order, so it will be important to assess how significant the omitted terms can be. It is likely that using the present kludge, augmented with conservative corrections up to 3.5PN order computed as outlined above, will do well enough at reproducing the phasing of true inspiral waveforms. As
discussed in \cite{pound}, it is likely that the conservative terms will be most important in the weak-field regime, where the post-Newtonian results are valid. In the regime where the post-Newtonian corrections cease to be valid, the contribution of conservative corrections to the phase evolution may be much less critical.  This will be investigated in future work. If the PN expansion is not sufficient, the kludge can be augmented using fits to the results of self-force calculations as described.


\section{Concluding discussion}
\label{conclusions}

In this paper we have provided a simple, ``easy-to-use'', prescription
for approximating the gravitational waveforms generated by test-bodies
inspiralling into Kerr black holes. These ``kludge'' waveforms are
constructed by combining familiar flat-spacetime wave equations with a
true geodesic trajectory in the Kerr spacetime. Despite its formal
inconsistency this hybrid approximation results in remarkably accurate
waveforms, which we have established by comparison to rigorous
Teukolsky-based perturbative waveforms. We find an impressive overlap
between the kludge and TB waveforms for particles on geodesic orbits
(i.e., ignoring radiation-reaction). This overlap is $>95 \%$ over a
significant portion of the relevant orbital parameter space.
Significant degradation (overlap reduced to $\sim 50\% $) occurs for
strong field orbits around rapidly spinning black holes. For such
cases, the contribution of the radiation backscattered from the
background spacetime is sizable; this effect is not included in the
present formulation of the kludge waveforms. As a rough (but reliable)
rule of thumb, we have found that kludge waveforms work well for Kerr
orbits with periaspe distance $r_p \gtrsim 5 M $, irrespective of the
black hole spin.

We have experimented with three different types of kludge waveforms
generated using three different solutions of the flat-spacetime
gravitational wave-equation: quadrupole, quadrupole-octupole and
``Press'' waveforms \cite{press}. The latter choice includes
contributions from all multipole moments of the orbiting body. We have
concluded that for all practical purposes the quadrupole-octupole
waveforms are the optimal choice, nearly as accurate as the Press
waveforms while much quicker and easier to generate.

Within the adiabatic approximation, we have applied our method to
calculate full inspiral kludge waveforms taking into account orbital
evolution. This realistic scenario requires an additional ``kludge''
for describing the orbital evolution itself. Our earlier work
\cite{GHK},\cite{GHK2} provides such a scheme: exact geodesic orbital
dynamics coupled with approximate PN-based expressions for the
fluxes. The resulting waveforms still have very high overlaps with
available Teukolsky-based inspirals for circular-inclined orbits. The
overlap is $\gtrsim 75 \% $ even when the inspiral terminates in the
strong field region. This result indicates the kludge will be a very
useful tool for generating inspiral waveforms. However, as yet
Teukolsky-based inspirals are available only for circular-inclined
orbits. Only when comparisons for generic inspiral orbits are possible
will we be able to firmly identify the range of applicability of these
kludge waveforms.

We have also used the kludge waveforms to estimate the fluxes of
energy and angular momentum carried away in GWs from geodesic
orbits. We have found that such kludge fluxes in general are not as
accurate for evolving inspirals as the post-Newtonian based
expressions described in \cite{GHK2}. The kludge fluxes do provide us
with an estimate of the error in using kludge waveforms for SNR
calculations, and our results suggest the kludge waveforms will tend
to underestimate the true SNRs if anything.

The area where these kludge waveforms will find most use is in the
development of EMRI data analysis for LISA. The combination of
accuracy and simplicity of generation has already made these waveforms
invaluable tools for the study of data analysis issues. It seems quite
plausible that the waveforms may also play a role in the final search
of the LISA data. One use could be for estimation of the waveform
parameters as the first stage in a hierarchical search. The high
faithfulness of these waveforms suggest that they may be able to set
fairly tight bounds on the parameters of the emitting system. This
will be extremely useful input for the second stage of the search
where the system parameters will be refined using more accurate
waveforms.  However, as we discussed before, the kludge waveforms at
present do not include some important physical features that we expect
in true inspirals.  Only when accurate Teukolsky or self-force based
waveforms are available for generic inspiral orbits will we be able to fully
quantify the range of validity and level of accuracy of the kludge
waveforms.

Time-frequency searches \cite{wengair05,gairwen05,gairjones06} could also be used to detect EMRIs in the first stage of a hierarchical search. These can typically detect EMRI events at about half the distance of the semi-coherent search \cite{rates}, without the use of templates. A template with overlap of $\sim50\%$ with the true signal can detect that signal at about half the distance of the correct template. The NK waveforms can easily achieve overlaps of $\sim50\%$ through most of parameter space, particularly given the $2-3$ week integration times needed for the semi-coherent approach. It is clear therefore that a first stage semi-coherent search using kludge waveforms would have a greater reach than time-frequency methods, and would put tighter constraints on the parameters. However, this would come at a much greater computational cost. The final LISA search will undoubtedly employ both methods at various stages in the analysis.

A further possible application of kludge waveforms is to the study of
non-Kerr EMRIs. It is hoped that LISA observations will allow
``spacetime-mapping'' of black holes \cite{ryan95}, \cite{KipRev}, and
thereby test the no-hair theorem. To carry out such tests
quantitatively will require waveform templates, which incorporate the
deviation from Kerr in the spacetime structure as a set of suitable
parameters (e.g., multipole moments \cite{ryan95}). The development of
rigorous non-Kerr EMRI waveforms is a very difficult task, since
generic spacetimes lack some symmetries that allowed the formulation
of the Teukolsky framework for the Kerr spacetime. One way to make
progress is to construct kludged waveforms in non-Kerr spacetimes
along the lines outlined in this paper. Our results for the Kerr
spacetime suggest that such kludged waveforms may well be sufficiently
accurate for qualitative, and even quantitative studies of non-Kerr
EMRIs. All that is required is the integration of geodesic equations
in the non-Kerr spacetime and thus kludge waveforms provide a
computationally quick and easy tool to study ``bumpy'' and
``quasi-Kerr'' spacetime mapping \cite{bumpy,qkerr}.

The kludged waveforms/inspirals presented in this and a companion
paper \cite{GHK2} could be considered as ``second generation'', an
improvement of the original simple version of quadrupole waveforms
\cite{webpage} and inspirals \cite{GHK}. Certainly, there is space for
further improvement. The inclusion of conservative self-force effects
on the inspiral is an obvious next step, and we have already
discussed here how this can be achieved by calculating the relevant
orbital frequency shifts for circular Schwarzschild orbits. We can
also further improve waveform generation by including the
back-scattering effect arising from the propagation of the GWs in a
curved spacetime.

In summary, given the present level of performance of our kludge
waveforms and inspirals and their prospects of improvement, we feel
that they should remain valuable tools for LISA source modelling for
the coming years.


\acknowledgments We thank Kip Thorne for initially pointing us to the
paper by Press, Steve Drasco for providing Teukolsky based waveforms
from generic orbits and B.S. Sathyaprakash for useful discussions. The
work of JRG was supported in part by NASA grants NAG5-12834 and
NAG5-10707 and by St.Catharine's College, Cambridge.  KG acknowledges
support from PPARC Grant PPA/G/S/2002/00038. Work of SB was partially
supported by PPARC Grant PP/B500731. Work of HF was supported in parts
by NASA grants NAG5-12834, NNG04GK98G and by NSF grants PHY-0099568
and PHY-0601459.  The work of SAH was supported by NASA Grants
NAG5-12906 and NNG05G105G, NSF Grant PHY-0244424 and CAREER grant
PHY-0449884.  SAH also gratefully acknowledges support from MIT's
Class of 1956 Career Development Fund.


\appendix
\section{Tables}

\begin{table}
\caption{\label{o-lap_equat}
Numerical data for overlaps between TB and kludge
waveforms \--- equatorial-eccentric Kerr orbits. Data is not shown with WD confusion noise for the ten orbits in the very strong field regime. In this regime, none of the kludge waveforms reproduce the TB waveforms very well, and this is compounded when the dominant harmonics are suppressed by white dwarf confusion noise. The overlaps are uniformly poor and uninformative, so we do not include them.}
\begin{ruledtabular}
\begin{tabular}{ccccc|ccc|ccc|c}
$p/M$ \footnotemark[1]&$e$\footnotemark[2] &$\iota$\footnotemark[3]
(deg)&$a/M$\footnotemark[4]
&$\Theta$\footnotemark[5] (deg)&\multicolumn{3}{c}{overlap($+$)\footnotemark[6]}
&\multicolumn{3}{c}{overlap with WD
($+$)}
&duration(M)\footnotemark[8] \\ & & & & &Quad&Quad-Oct&Press&Quad&Quad-Oct&Press& \\
\hline
1.7 &  0.1&   0& 0.99&  90& 0.84&0.772&0.741  & & & & 2000 \\
1.7 &  0.3&   0& 0.99&  90& 0.76&0.557&0.500  & & & & 700  \\
1.9 &  0.5&   0& 0.99&  90& 0.544&0.570&0.547  & & & & 700  \\
1.9 &  0.5&   0& 0.99&  90& 0.523&0.484&0.445  & & & & 2000 \\
2.11&  0.7&   0& 0.99&  45&  0.562&0.566&0.562  & & & & 700  \\
2.2 &  0.7&   0& 0.99&  90& 0.526&0.496&0.458  & & & & 700  \\
2.5 &  0.1&   0& 0.99&  90& 0.906&0.853&0.827  & & & & 2000 \\
2.5 &  0.5&   0& 0.99&  90& 0.671&0.665&0.651  & & & & 700  \\
3.5 &  0.4&   0& 0.99&  90& 0.588&0.524&0.507  & & & & 2000 \\
3.5 &  0.4&   0& 0.99&  45& 0.624&0.598&0.593  & & & & 5000 \\
5.1 &  0.5&   0& 0.5 &  90& 0.856&0.962&0.967  &0.856&0.961&0.968 & 700  \\
5.5 &  0.5&   0& 0.5 &  90& 0.864&0.964&0.973  &0.862&0.962&0.973 & 2000 \\
6.0 &  0.4&   0& 0.5 &  90& 0.871&0.970&0.980  &0.864&0.967&0.979 & 2000 \\
6.0 &  0.5&   0& 0.5 &  90& 0.858&0.966&0.974  &0.855&0.963&0.973 & 2000 \\
6.5 &  0.5&   0& 0.5 &  90& 0.870&0.970&0.979  &0.864&0.968&0.978 & 2000 \\
6.5 &  0.5&   0& 0.5 &  45& 0.937&0.987&0.990  &0.932&0.986&0.990 & 8000 \\
10.0&  0.3& 180& 0.99&  90& 0.864&0.961&0.966  &0.806&0.943&0.954 & 8000 \\
10.0&  0.3& 180& 0.99&  45& 0.922&0.971&0.969  &0.883&0.957&0.956 & 8000 \\
10.4&  0.5& 180& 0.99&   0& 0.998&0.998&0.999  &0.997&0.999&0.998 &2000 \\
10.5&  0.5& 180& 0.99&  90& 0.878&0.975&0.982  &0.856&0.968&0.978 & 2000 \\
15.0&  0.4&   0& 0.5 &  90& 0.824&0.963&0.968  &0.60&0.878&0.881 &  8000 \\
15.0&  0.4&   0& 0.99&  90& 0.824&0.961&0.963  &0.603&0.874&0.860 & 8000 \\
\end{tabular}
\end{ruledtabular}
\footnotetext[1]{semi-latus rectum}
\footnotetext[2] {eccentricity}
\footnotetext[3] {inclination angle}
\footnotetext[4]{spin}
\footnotetext[5]{observation point, $\phi=0$ always}
\footnotetext[6]{overlap between ``$+$'' polarization of TB waveform with quadrupole (``Quad''), quadrupole-octupole (``Quad-Oct'') and Press (``Press'') kludge waveforms}
\footnotetext[7]{overlap between ``$\times$'' polarizations}
\footnotetext[8] {waveform duration}
\end{table}

\newpage

\begin{table}
\caption{\label{o-lap_circ}
Numerical data for overlaps between TB and kludge
waveforms \--- inclined-circular Kerr orbits.}
\begin{ruledtabular}
\begin{tabular}{ccccc|ccc|ccc|ccc|c}
$p/M$ &$e$ &$\iota$ (deg)&$a/M$ &$\Theta$ (deg)&\multicolumn{3}{c}{overlap($+$)}
&\multicolumn{3}{c}{overlap($\times$)}& \multicolumn{3}{c}{overlap with WD ($\times$)} & duration(M) \\
& & & & &Quad&Quad-Oct&Press&Quad&Quad-Oct&Press&Quad&Quad-Oct&Press& \\
\hline
5.0 &  0&  30& 0.5 &  45& 0.944&0.984&0.990  &0.946&0.984&0.990  &0.945&0.984&0.990 &3000 \\
5.0 &  0&  30& 0.5 &  90& 0.899&0.974&0.984  &0.891&0.971&0.984  &0.89 &0.969&0.983 &3000 \\
5.0 &  0&  30& 0.99&  45& 0.929&0.969&0.975  &0.93 &0.969&0.975  &0.927&0.967&0.974 &3000 \\
5.0 &  0&  30& 0.99&  90& 0.904&0.958&0.964  &0.9  &0.957&0.966  &0.895&0.954&0.965 &3000 \\
5.05&  0&  60& 0.5 &  45& 0.924&0.967&0.973  &0.922&0.969&0.977  &0.922&0.968&0.976 &3000 \\
5.05&  0&  60& 0.5 &  90& 0.91 &0.955&0.961  &0.911&0.963&0.973  &0.907&0.961&0.972 &3000 \\
5.0 &  0&  60& 0.99&  45& 0.857&0.912&0.917  &0.86 &0.917&0.925  &0.857&0.913&0.921 &3000 \\
5.0 &  0&  60& 0.99&  90& 0.854&0.882&0.888  &0.87 &0.912&0.925  &0.862&0.907&0.922 &3000 \\
\hline
10.0&  0&  30& 0.5 &  45& 0.93 &0.989&0.995  &0.936&0.99 &0.994  &0.751&0.957&0.975 &8000 \\
10.0&  0&  30& 0.5 &  90& 0.89 &0.981&0.990  &0.901&0.981&0.990  &0.63 &0.93 &0.969 &7000 \\
10.0&  0&  30& 0.99&  45& 0.92 &0.98 &0.986  &0.93 &0.98 &0.986  &0.674& 0.92&0.946 &8000 \\
10.0&  0&  30& 0.99&  90& 0.884&0.974&0.982  &0.915&0.975&0.983  &0.371&0.892&0.963 &7000 \\
10.0&  0&  45& 0.7 &  45& 0.932&0.980&0.981  &0.936&0.983&0.987  &0.75 &0.935&0.953 &8000 \\
10.0&  0&  45& 0.7 &  90& 0.922&0.972&0.972  &0.92 &0.98 &0.987  &0.68 &0.929&0.969 &8000 \\
10.0&  0&  60& 0.5 &  45& 0.914&0.981&0.987  &0.912&0.982&0.990  &0.691&0.937&0.969 &8000 \\
10.0&  0&  60& 0.5 &  90& 0.945&0.98 &0.981  &0.932&0.981&0.986  &0.732&0.925&0.948 &7000 \\
10.0&  0&  60& 0.99&  45& 0.873&0.953&0.958  &0.875&0.957&0.965  &0.611&0.875&0.906 &8000 \\
10.0&  0&  60& 0.99&  90& 0.923&0.941&0.937  &0.915&0.959&0.965  &0.675&0.86 &0.893 &7000 \\
\hline
20.0&  0&  30& 0.5 &  45& 0.934&0.987&0.982  &0.938&0.989&0.989  &0.957&0.987&0.902 &50000\\
20.0&  0&  30& 0.5 &  90& 0.889&0.970&0.922  &0.893&0.974&0.962  &0.92 &0.984&0.900 &30000\\
20.0&  0&  30& 0.99&  45& 0.93 &0.98 &0.951  &0.94 &0.988&0.992  &0.951&0.987&0.977 &50000\\
20.0&  0&  30& 0.99&  90& 0.895&0.968&0.907  &0.914&0.979&0.988  &0.933&0.979&0.951 &30000\\
20.0&  0&  60& 0.5 &  45& 0.92 &0.975&0.966  &0.916&0.981&0.995  &0.952&0.958&0.990 &50000\\
20.0&  0&  60& 0.5 &  90& 0.95 &0.973&0.958  &0.934&0.976&0.972  &0.95 &0.967&0.957 &30000\\
20.0&  0&  60& 0.99&  45& 0.895&0.963&0.962  &0.892&0.967&0.973  &0.934&0.951&0.896 &50000\\
20.0&  0&  60& 0.99&  90& 0.954&0.96 &0.938  &0.937&0.972&0.978  &0.963&0.946&0.943 &30000\\
\end{tabular}
\end{ruledtabular}
\end{table}
\newpage

\begin{table}
\caption{\label{o-lap_generic}
Numerical data for overlaps between TB and kludge
waveforms \--- generic Kerr orbits.}
\begin{ruledtabular}
\begin{tabular}{ccccc|ccc|ccc|ccc|c}
$p/M$ &$e$ &$\iota$ (deg)&$a/M$ &$\Theta$ (deg)&\multicolumn{3}{c}{overlap($+$)}
&\multicolumn{3}{c}{overlap($\times$)}& \multicolumn{3}{c}{overlap with WD ($\times$)} & duration(M) \\
& & & & &Quad&Quad-Oct&Press&Quad&Quad-Oct&Press&Quad&Quad-Oct&Press& \\
\hline
6.0 &  0.1&  20.1364& 0.9 &  90& 0.912&0.982&0.993  &0.91 &0.984&0.996   &0.894&0.98 &0.995  &15000 \\
6.0 &  0.1&  20.1364& 0.9 &  60& 0.935&0.988&0.996  &0.939&0.99 &0.997   &0.929&0.989&0.997  &15000\\
6.0 &  0.1&  20.1364& 0.9 &  30& 0.972&0.995&0.998  &0.972&0.995&0.998   &0.968&0.995&0.998  &15000\\
\hline
6.0 &  0.5&  20.103&  0.9 &  90& 0.89 &0.967&0.973  &0.89 &0.972&0.980   &0.875&0.968&0.979  &15000\\
6.0 &  0.5&  20.103&  0.9 &  60& 0.916&0.975&0.980  &0.912&0.978&0.982   &0.913&0.976&0.981  &15000\\
6.0 &  0.5&  20.103&  0.9 &  30& 0.958&0.985&0.985  &0.959&0.985&0.986   &0.955&0.984&0.985  &15000\\
\hline
6.0 &  0.1&  60.1461& 0.9 &  90& 0.951&0.989&0.993  &0.941&0.988&0.995   &0.933&0.987&0.995  &10000\\
6.0 &  0.1&  60.1461& 0.9 &  60& 0.958&0.989&0.994  &0.95 &0.988&0.995   &0.946&0.987&0.996  &10000\\
6.0 &  0.1&  60.1461& 0.9 &  30& 0.943&0.986&0.996  &0.942&0.987&0.996   &0.941&0.986&0.996  &10000\\
\hline
6.0 &  0.5&  60.1108& 0.9 &  90& 0.934&0.975&0.980  &0.919&0.974&0.982   &0.912&0.971&0.981  &15000\\
6.0 &  0.5&  60.1108& 0.9 &  60& 0.939&0.974&0.982  &0.927&0.972&0.984   &0.924&0.97 &0.984  &15000\\
6.0 &  0.5&  60.1108& 0.9 &  30& 0.917&0.971&0.978  &0.916&0.971&0.979   &0.915&0.971&0.979  &15000\\
\hline
6.0 &  0.7&  60.0755& 0.9 &  90& 0.926&0.966&0.970  &0.911&0.967&0.974   &0.906&0.966&0.975  &20000\\
6.0 &  0.7&  60.0755& 0.9 &  60& 0.928&0.967&0.971  &0.919&0.966&0.974   &0.916&0.966&0.974  &20000\\
6.0 &  0.7&  60.0755& 0.9 &  30& 0.897&0.963&0.966  &0.897&0.963&0.967   &0.897&0.963&0.968  &20000\\
\hline
12.0&  0.1& 119.9586& 0.9 &  90& 0.916&0.987&0.994  &0.911&0.988&0.998   &0.627&0.922&0.985  &15000\\
12.0&  0.1& 119.9586& 0.9 &  60& 0.926&0.988&0.991  &0.92 &0.989&0.998   &0.634&0.926&0.985  &15000\\
12.0&  0.1& 119.9586& 0.9 &  30& 0.9  &0.983&0.994  &0.899&0.986&0.994   &0.577&0.91 &0.956  &15000\\
\hline
12.0&  0.5& 119.9686& 0.9 &  90& 0.938&0.988&0.992  &0.934&0.988&0.995   &0.889&0.979&0.993  &15000\\
12.0&  0.5& 119.9686& 0.9 &  60& 0.942&0.988&0.993  &0.936&0.987&0.994   &0.889&0.977&0.991  &15000\\
12.0&  0.5& 119.9686& 0.9 &  30& 0.924&0.984&0.993  &0.924&0.986&0.994   &0.862&0.971&0.988  &15000\\
\hline
12.0&  0.7& 119.9786& 0.9 &  90& 0.935&0.986&0.992  &0.933&0.987&0.993   &0.904&0.981&0.991  &25000\\
12.0&  0.7& 119.9786& 0.9 &  60& 0.940&0.987&0.992  &0.934&0.985&0.992   &0.904&0.978&0.990  &25000\\
12.0&  0.7& 119.9786& 0.9 &  30& 0.926&0.983&0.992  &0.924&0.983&0.990   &0.884&0.973&0.986  &25000\\
\hline
12.0&  0.3& 139.9597& 0.9 &  90& 0.9  &0.985&0.992  &0.891&0.984&0.996   &0.751&0.955&0.991  &15000\\
12.0&  0.3& 139.9597& 0.9 &  60& 0.938&0.991&0.994  &0.931&0.99 &0.997   &0.827&0.972&0.993  &15000\\
12.0&  0.3& 139.9597& 0.9 &  30& 0.948&0.992&0.994  &0.949&0.993&0.998   &0.859&0.979&0.994  &15000\\
\hline
12.0&  0.1& 159.9728& 0.9 &  90& 0.836&0.973&0.992  &0.838&0.974&0.993   &0.494&0.887&0.968  &15000\\
12.0&  0.1& 159.9728& 0.9 &  60& 0.881&0.984&0.995  &0.889&0.987&0.998   &0.549&0.926&0.990  &15000\\
12.0&  0.1& 159.9728& 0.9 &  30& 0.947&0.994&0.997  &0.947&0.995&0.997   &0.726&0.96 &0.968  &15000\\
\hline
12.0&  0.5& 159.9793& 0.9 &  90& 0.876&0.976&0.991  &0.877&0.977&0.995   &0.808&0.961&0.993  &15000\\
12.0&  0.5& 159.9793& 0.9 &  60& 0.912&0.985&0.994  &0.919&0.988&0.996   &0.862&0.977&0.994  &15000\\
12.0&  0.5& 159.9793& 0.9 &  30& 0.961&0.994&0.996  &0.962&0.995&0.998   &0.932&0.991&0.996  &15000\\
\end{tabular}
\end{ruledtabular}
\end{table}

\begin{table}
\begin{ruledtabular}
\begin{tabular}{cccccccccc}
$a/M$ & $p/M$ &  $e$  &  $\iota$   & $\langle \dot{E} \rangle_{TB}$ & $ \langle \dot{E} \rangle_{GG}$ &
$ \langle \dot{E} \rangle_{quad}$ &
$\langle \dot{L}_z \rangle_{TB}$ & $ \langle \dot{L}_z \rangle_{GG}$ &  $ \langle \dot{L}_z \rangle_{quad}$ \\
\hline
0.95 & 100 & 0 & 60. & $-6.219e-10$ & $-6.219e-10$ & $-6.381$e-10 &
$-3.118$e-7& $-3.122$e-7 & $-3.204$e-7 \\
\hline
0.5 & 5 & 0.3 & 0. & $-2.604$e-3 & $-2.17$e-3 & $-2.08$e-3& $-2.48$e-2&
$-2.00$e-2& $-2.03$e-2 \\
0.5 & 5 & 0.4 & 0. & $-3.530$e-3 & $-2.37$e-3 & $-2.57$e-3& $-2.98$e-2&
$-1.94$e-2 & $-2.24$e-2  \\
0.5 & 6 & 0.3 & 0. & $-8.883$e-4 & $-8.45$e-4 & $-7.99$e-4& $-1.13$e-2&
$-1.03$e-2& $-1.03$e-2 \\
0.5 & 6 & 0.4 & 0. & $-1.033$e-3 & $-9.17$e-4 & $-8.77$e-4& $-1.18$e-2&
$-1.00$e-2 & $-1.03$e-2 \\
0.5 & 6 & 0.5 & 0. & $-1.196$e-3 & $-9.51$e-4 & $-9.50$e-4& $-1.23$e-2&
$-9.41$e-3 &
$-1.00$e-2 \\
0.9 & 12.152 & 0.3731 & 0. & $-2.357$e-5 & $-2.352$e-5 & $-2.47$e-5&
 $8.1351$e-4 & $-7.99$e-4& $-8.46$e-4 \\
0.99 & 2 & 0.1 & 0. & $-4.723$e-2 & $-3.94$e-2 & $-3.34$e-2& $-0.178$&
$-0.120$& $-0.126$ \\
0.99 & 2 & 0.3 & 0. & $-5.634$e-2 & $-5.084$e-2 & $-3.62$e-2& $-0.190$&
$2.39$e-3 & $-0.125$ \\
0.99 & 2 & 0.4 & 0. & $-6.429$e-2 & $-4.11$e-2 & $-3.90$e-2& $-0.201$&
$9.55$e-2 & $-0.125$ \\
0.99 & 3 & 0.1 & 0. & $-1.124$e-2 & $-1.096$e-2 & $-9.33$e-3&
$-6.83$e-2&
$-6.29$e-2 & $-5.69$e-2 \\
0.99 & 3 & 0.3 & 0. & $-1.315$e-2 & $-1.434$e-2 & $-9.88$e-3&
$-7.10$e-2&
$-5.47$e-2& $-5.46$e-2 \\
0.99 & 11 & 0.2 & 180. & $-5.639$e-5 & $-5.25$e-5 & $-5.36$e-5&
$1.81$e-3&
$1.71$e-3& $1.73$e-3 \\
0.99 & 11 & 0.4 & 180. & $-8.337$e-5 & $-6.12$e-5 & $-7.25$e-5&
$2.13$e-3&
$1.65$e-3& $1.89$e-3 \\
0.99 & 11 & 0.5 & 180. & $-1.049$e-4 & $-6.38$e-5 & $-8.60$e-5&
$2.36$e-3& $1.55$e-3& $1.98$e-3 \\
\hline
0.9 & 6 & 0.1 & 40.192285 & -6.1850e-4  &  -6.196e-4  & -6.112e-4 & -7.628e-3  & -7.534e-3 & -7.547e-3 \\
0.9 &6 &    0.3 &   40.176668 &     -7.2678e-4  &  -7.512e-4  & -6.744e-5 & -7.835e-3  & -7.632e-3 & -7.359e-3  \\
0.9 &6 &    0.5 &   40.145475 &     -8.7445e-4  &  -8.475e-4  & -7.344e-4 & -7.8084e-3 & -7.143e-3 & -6.686e-3  \\
0.9 &6 &    0.7 &   40.098788 &     -8.7037e-4  &  -6.834e-4  & -6.506e-4 & -6.488e-3  & -5.140e-3 & -4.945e-3  \\
0.9 &6 &    0.1 &   80.046323 &     -8.0701e-4  &  -8.007e-4  & -7.283e-4 & -3.6182e-3 & -3.395e-3 & -3.328e-3  \\
0.9 &6 &    0.3 &   80.042690 &     -1.0863e-3  &  -9.879e-4  & -8.851e-4 & -4.369e-3  & -3.659e-3 & -3.698e-3  \\
0.9 &6 &    0.5 &   80.035363 &     -1.6792e-3  &  -1.163e-3  & -1.167e-3 & -5.874e-3  & -3.761e-3 & -4.321e-3  \\
0.9 &6 &    0.7 &   80.024226 &     -2.6923e-3  &  -1.010e-3  & -1.537e-3 & -8.251e-3  & -3.002e-3 & -5.056e-3  \\
0.9 &12&    0.1 &   99.982344 &     -2.5084e-5  &  -2.511e-5  & -2.621e-5 &  1.228e-3  &  1.238e-4 & 1.261e-4   \\
0.9 &12&    0.3 &   99.983766 &     -2.9241e-5  &  -2.900e-5  & -2.963e-5 &  1.143e-4  &  1.150e-4 & 1.143e-4   \\
0.9 &12&    0.5 &   99.986612 &     -3.4405e-5  &  -3.231e-5  & -3.315e-5 &  9.515e-5  &  9.419e-5 & 8.989e-5   \\
0.9 &12&    0.7 &   99.990888 &     -3.2914e-5  &  -2.749e-5  & -2.973e-5 &  6.196e-5  &  5.865e-5 & 5.345e-5   \\
0.9 &12&    0.1 &   139.956153 & -2.8394e-5  &  -2.828e-5  & -2.883e-5 &  8.406e-4  &  8.389e-4 & 8.530e-4  \\
0.9 &12&    0.3 &   139.959664 &    -3.445e-5   &  -3.265e-5  & -3.361e-5 &  8.6536e-4  & 8.265e-4  & 8.481e-4  \\
0.9 &12&    0.5 &   139.966702 &    -4.3778e-5  &  -3.657e-5  & -3.963e-5 &  8.674e-4  &  7.506e-4  & 7.964e-4  \\
0.9 &12&    0.7 &   139.977303 &    -4.623e-5  &  -3.153e-5  & -3.818e-5  &  7.287e-4  &  5.343e-4  & 6.121e-4  \\
\end{tabular}
\end{ruledtabular}
\caption[Table]{A comparison of the approximate fluxes computed from
the kludge quadrupole waveforms \erf{quad} described in this paper to
accurate Teukolsky based fluxes for circular (top section), equatorial
(middle section) and generic (bottom section) orbits. The energy fluxes
are expressed in units of $\mu^2/M$, and the angular momentum fluxes in
units of $\mu^2/M^2$. The subscripts ``TB'', ``GG'' and ``quad'' denote
fluxes computed from Teukolsky based waveforms, from the kludge flux
expressions in \cite{GHK2} and the approximate quadrupole fluxes
respectively.}
\label{tab3}
\end{table}

\begin{table}
\begin{ruledtabular}
\begin{tabular}{cccccccccc}
$a/M$ & $p/M$  & $e$  &  $\iota $ & $\Delta E_{TB}$ & $\Delta E_{GG}$ &  $\Delta E_{quad}$ & $\Delta {Lz}_{TB}$ & $\Delta {Lz}_{GG}$ &
$\Delta {Lz}_{quad}$\\
\hline
0 & 8.001 & 1 &  0 &  $-2.28$ & $-0.157$ & $-0.976$ & $-19.1$ & $-0.664$ & $-8.45$ \\
0 & 10  & 1 &  0 & $-1.47$e-1 & $-6.62$e-2 & $-0.101$ & $-1.90$ &
$-0.971$ & $-1.36$ \\
0 & 14  & 1 &  0 & $-2.93$e-2 & $-1.87$e-2 & $-2.44$e-2 &
$-0.631$ &
$-0.450$ &
$-0.540$ \\
0 & 20  & 1 &  0 & $-6.78$e-3 & $-5.08$e-3 & $-6.26$e-3 &
$-0.250$ &
$-0.206$ &
$-0.235$ \\
0 & 50  & 1 &  0 & $-2.199$e-4 & $-1.97$e-4 & $-2.21$e-4 &
$-3.26$e-2 &
$-3.05$e-2 &
$-3.28$e-2 \\
\hline
0.9 & 15    & 1     &   0  &  \--- & -0.01115  &  -0.01441 & \--- &  -0.3190    &  -0.3646\\
0.9 & 15  & 1 &  60  & \--- &  -0.01284  & -0.01566 & \--- &   -0.1999  &   -0.2204 \\
0.9 & 15  & 1 &  120 & \--- &  -0.01633  & -0.02057 & \--- &   0.1882  &  0.2126 \\
0.9 & 15  & 1 &  180 & \--- &  -0.01815  & -0.02527 & \--- &  0.4656  &  0.6035 \\
\hline
0.99 & 15 & 1 &   0   & \--- & -0.01084 & -0.01403 & \--- & -0.3129  &    -0.3567 \\
0.99 & 15 & 1 &  60   & \--- & -0.01269 & -0.01531 & \--- & -0.2007  &  -0.2184 \\
0.99 & 15 & 1 &  120  & \--- & -0.01653 & -0.02064 & \--- & 0.1878   &   0.2097 \\
0.99 & 15 & 1 &  180  & \--- & -0.01853 & -0.02626 & \--- & 0.4744   & 0.6250 \\
\end{tabular}
\end{ruledtabular}
\caption[Table]{As in Table~\ref{tab3}, but for parabolic orbits.}
\label{tab4}
\end{table}


\end{document}